\documentclass[aps,prx,twocolumn,superscriptaddress,amsmath,epsfig,amssymb]{revtex4}
\usepackage{makecell}
\usepackage{bbm}
\usepackage{mathrsfs}
\usepackage{amsfonts}
\usepackage{color}
\usepackage{graphicx}
\newcommand{\dbar}{d\hspace*{-0.08em}\bar{}\hspace*{0.1em}}
\newcommand{\mec}[1]{{\color{red} #1}}
\begin{document}
\newcommand{\Rv}{\boldsymbol R}
\newcommand{\rv}{{\boldsymbol r}}
\newcommand{\tv}{{\vec t}}
\newcommand{\av}{\boldsymbol a}
\newcommand{\fv}{{\boldsymbol f}}
\newcommand{\mv}{{\boldsymbol m}}
\newcommand{\nv}{{\boldsymbol n}}
\newcommand{\hv}{{\boldsymbol h}}
\newcommand{\xv}{{\boldsymbol x}}
\newcommand{\xp}{\vec{x}_{\perp}}
\newcommand{\pp}{\vec{p}_{\perp}}
\newcommand{\zv}{{\boldsymbol z}}
\newcommand{\sv}{{\boldsymbol s}}
\newcommand{\uv}{{\vec u}}
\newcommand{\Av}{{\boldsymbol A}}
\newcommand{\Fv}{{\boldsymbol F}}
\newcommand{\Xv}{{\boldsymbol X}}
\newcommand{\Yv}{{\boldsymbol Y}}
\newcommand{\Pv}{{\boldsymbol P}}
\newcommand{\Qv}{{\boldsymbol Q}}
\newcommand{\Hv}{{\boldsymbol H}}
\newcommand{\ur}{\vec{{\EuFrak u}}}
\newcommand{\cv}{{\vec c}}
\newcommand{\qv}{{\boldsymbol q}}
\newcommand{\pv}{{\boldsymbol p}}
\newcommand{\wv}{{\boldsymbol w}}
\newcommand{\yv}{{\boldsymbol y}}
\newcommand{\vv}{{\boldsymbol v}}
\newcommand{\kv}{{\vec k}}
\newcommand{\phiv}{{\boldsymbol \phi}}
\newcommand{\etav}{{\boldsymbol \eta}}
\newcommand{\Tr}{{\rm Tr}}
\newcommand{\px}{{\partial_x}}
\newcommand{\py}{{\partial_y}}
\newcommand{\ppi}{{\partial_i}}
\newcommand{\ppj}{{\partial_j}}
\newcommand{\ch}{{\hat{c}}}
\newcommand{\eh}{{\hat{e}}}
\newcommand{\xh}{{\hat{x}}}
\newcommand{\yh}{{\hat{y}}}
\newcommand{\zh}{{\hat{z}}}
\newcommand{\vh}{{\hat{v}}}
\newcommand{\qh}{{\hat{q}}}
\newcommand{\kh}{{\hat{k}}}
\newcommand{\llm}{{\boldsymbol \lambda}}
\newcommand{\Am}{{\boldsymbol{A}}}
\newcommand{\Qm}{{\boldsymbol{Q}}}
\newcommand{\Rm}{{\boldsymbol R}}
\newcommand{\Lm}{{\boldsymbol L}}
\newcommand{\Km}{{\boldsymbol K}}
\newcommand{\Jm}{{\boldsymbol J}}
\newcommand{\Tm}{{\boldsymbol T}}
\newcommand{\Bm}{{\boldsymbol B}}
\newcommand{\Dm}{{\boldsymbol D}}
\newcommand{\Cm}{{\boldsymbol C}}
\newcommand{\Em}{{\boldsymbol E}}
\newcommand{\Mm}{{\mathcal M}}
\newcommand{\Wm}{{\boldsymbol W}}
\newcommand{\Fm}{{\boldsymbol F}}
\newcommand{\Gm}{{\boldsymbol G}}
\newcommand{\Imm}{{\boldsymbol I}}
\newcommand{\sm}{{\boldsymbol s}}
\newcommand{\gammam}{{\boldsymbol \Gamma}}
\newcommand{\chim}{{\boldsymbol \chi}}
\newcommand{\be}{\ba}
\newcommand{\ee}{\ea}
\newcommand{\ba}{\begin{eqnarray}}
\newcommand{\ea}{\end{eqnarray}}
\newcommand{\RNum}[1]{\uppercase\expandafter{\romannumeral #1\relax}}
\newcommand{\ddelta}{\boldsymbol{\delta}}
\newcommand{\thetav}{\boldsymbol{\theta}}
\newcommand{\pdf}{\mathcal{P}}
\newcommand{\U}{\mathcal{U}}
\newcommand{\LFP}{\mathcal{L}_{\rm FP}}

\title{Microscopic Dynamical Entropy\\ I:
Quantifying Hamiltonian Irreversibility in Large and Small Systems}
\author{Mingnan Ding}
\email{dmnphy0@gmail.com}
\affiliation{Wilczek Quantum Center, School of Physics and Astronomy, Shanghai Jiao Tong University, Shanghai 200240, China}
\affiliation{DAMTP, Centre for Mathematical Sciences, University of Cambridge, Wilberforce Road, Cambridge CB3 0WA, United Kingdom}

\author{Michael E. Cates}
\affiliation{DAMTP, Centre for Mathematical Sciences, University of Cambridge, Wilberforce Road, Cambridge CB3 0WA, United Kingdom}

\date{\today} 

\begin{abstract}

We introduce a Microscopic Dynamical Entropy (MDE) for Hamiltonian systems, defined with respect to a chosen partition of degrees of freedom into a system X and its environment Y.  The construction is based on the conditional phase-space volume (CPV), or equivalently the conditional Boltzmann entropy, associated with the 
unmonitored degrees of freedom Y.
Our MDE is a microscopically defined entropy functional of the marginal distribution $\rho_X(t)$, obtained by systematically discarding the conditional microscopic information associated with Y from the Gibbs entropy of the joint XY system. (For this joint system, exact Hamiltonian dynamics is retained at all times.)  
This construction clarifies the microscopic origins of thermal entropy in several ways.  
The dependence of MDE solely on $\rho_X(t)$ is consistent with the thermodynamic assumption that the entropy increment of a heat bath Y depends on its heat content and temperature only, not on details of its probability distribution. 
Indeed, the MDE fully recovers the familiar result $dS = \dbar Q/T$ connecting entropy increments to heat flow between system and environment. More generally, the MDE provides a consistent description of irreversible relaxation under exact Hamiltonian dynamics, while permitting transient entropy decreases to arise, as they do both in small systems, and in systems undergoing spin-echo type protocols. In systems with  time-scale separation between X and Y, the MDE becomes strictly monotonic in time, recovering quantitatively the familiar structure of irreversible thermodynamics. Remarkably, we find that the MDE can foreshadow thermodynamics even in a small isolated Hamiltonian system, if X is a well chosen subset of its degrees of freedom.  An example is the centre of mass of a set of interacting particles confined to a box. Here, even for as few as $N=10$ particles, the MDE increases during relaxation towards a maximum at equilibrium, and does so with increasing monotonicity at larger $N$.  Taken together, our results show that the MDE offers a powerful and fully consistent microscopic interpretation of the nonequilibrium thermal entropy and its time dependence, within the framework of exact Hamiltonian dynamics.
\end{abstract}

\maketitle

\section{Introduction}

Entropy lies at the heart of both thermodynamics and statistical mechanics. Despite its central role, however, a universally applicable definition of entropy is still lacking. This conceptual gap underlies the long-standing debates on the precise meaning and scope of the second law of thermodynamics~\cite{Cates2015,teVrugt2021,Jaynes1965,Callender2021,Lebowitz2005,xing2025}. Historically, the notion of entropy originated from the study of heat engines, where entropy is introduced as a thermodynamic quantity characterizing heat exchange~\cite{Carnot1872,Clausius1854,Clausius1865,Thomson1874}. In this framework, the thermal entropy itself is defined operationally, with only its differential $dS=\dbar Q/T$ specified through the interaction with an ideal heat bath. Boltzmann and Gibbs entropies were later proposed to provide microscopic counterparts~\cite{Boltzmann2012,Ehrenfest1959,Carlo2006,Gibbs1902}. Both are successful in equilibrium statistical mechanics, but neither works well out of equilibrium.

In particular, the Gibbs entropy is strictly conserved under exact Hamiltonian dynamics~\cite{Gibbs1902}. As a consequence, it cannot account for irreversible relaxation without invoking additional assumptions~\cite{Maes2003,Brush1966,Wu1975,Uffink2001,Uffink2007,Callender2021,Pippard1966,Gell-Mann1996}. Typically, procedures to coarse-grain phase-space cells have been introduced, creating monotonic entropy growth~\cite{Ehrenfest1959}, but such an approach is not conceptually satisfying as it alters the actual statistics of dynamical trajectories from that predicted by Hamiltonian dynamics via Liouville's equation. Furthermore, spin-echo experiments demonstrate that a genuinely microscopic entropy cannot be required to increase monotonically~\cite{Ridderbos1998}. More generally, as was known to both Maxwell and Kelvin~\cite{Thomson1874}, entropy decrease is not only possible but physically realizable at the microscopic level. (This is fully quantified in modern stochastic thermodynamics~\cite{Seifert2012}.) 

The Gibbs entropy provides a precise measure of the information content of a phase-space distribution. This property, however, also makes it conceptually distinct from thermal entropy. In thermodynamics, the entropy of a heat bath characterizes its macroscopic thermodynamic role and is largely insensitive to microscopic details of its internal state. By contrast, the Gibbs entropy of the bath depends explicitly on the full probability distribution over its microscopic degrees of freedom.

From this perspective, the Boltzmann entropy is conceptually closer to thermal entropy, as it characterizes the volume of dynamically accessible microscopic states compatible with given macroscopic constraints, rather than detailed statistical information~\cite{Boltzmann2012,Ehrenfest1959,Carlo2006,Lebowitz1999,Lebowitz2005,Gross2001}. This distinction highlights an essential asymmetry in statistical mechanics: the system is described in terms of its state or phase-space distribution and its time evolution, while the bath enters only through coarse (and generally constant) thermodynamic properties such as temperature. Indeed such an asymmetry is not only a cornerstone of classical thermodynamics, but explicitly built into the foundations of modern stochastic thermodynamics: in both cases the bath is treated as an effective environment rather than a fully resolved dynamical system~\cite{xing2025,Seifert2012}.

Following these considerations, we introduce the \emph{Microscopic Dynamical Entropy} (MDE). This is defined for an arbitrary Hamiltonian system in which 
we partition the degrees of freedom into two sets, X and Y, without assuming weak coupling or stochastic dynamics, and treat Y as the environment of X.  
The MDE is defined as the Gibbs entropy of the reduced phase-space density $\rho_X$ describing the state of X, supplemented by a contribution determined by the microscopic phase-space geometry of the eliminated degrees of freedom Y, conditioned on the state of X.  
Explicitly, the MDE consists of the sum of the Gibbs entropy of X and the average conditional Boltzmann entropy of Y.

We define the conditional Boltzmann entropy of Y through the \emph{conditional phase-space volume} (CPV),
$\Omega_Y(\xv,E)=\int d\yv\,\delta\!\left(E-H_{XY}(\xv,\yv)\right)$, 
which measures the accessible phase-space volume of Y at fixed total energy $E$ and fixed configuration $\xv$ of X. Here $H_{XY}$ is the Hamiltonian of the complete XY system.
Importantly, the CPV provides a well-defined reference measure on the state space of X, and determines the equilibrium distribution of X without invoking any ensemble assumption.  

The MDE can be viewed as the Gibbs entropy of the total system, {\em after} the conditional distribution $\rho_{Y|X}$ of `the environment', Y, has been replaced by the uniform distribution $\delta(E-H_{XY})/\Omega_Y(\xv,E)$ across $\yv\in$ Y. This `flattening' of $\rho_{Y|X}$ can be viewed as a  {\em non-destructive} coarse graining.  By non-destructive, we mean that  the exact Hamiltonian evolution of the marginal distribution $\rho_X$ is preserved. Note however that this interpretation is optional: the MDE is defined directly from the exact reduced state $\rho_X$ and the Hamiltonian geometry encoded in $\Omega_Y$, without necessarily appealing to coarse graining. 

Thus our construction of the MDE contrasts with a standard approach to defining a microscopic entropy, dating back to Gibbs himself, unavoidably based on {\em destructive} coarse-graining. By this we mean coarse-graining that {\em approximates the dynamics} rather than treating it exactly. Typically, phase space is partitioned into cells, within which $\rho$ is repeatedly replaced by its cell-wise average $\bar\rho$~\cite{Gibbs1902,Ehrenfest1959,Maes2003,teVrugt2021}. Because $S^G[\bar\rho]\ge S^G[\rho]$ for any function $\rho$, the coarse-grained Gibbs entropy $S^{\rm cg}(t) = S^G[\bar \rho]$ is strictly non-decreasing in time. 

There are two problems with the destructive approach, both well known. First, without a clear prescription on how cells are chosen, it offers no {\em quantitative} theory of heat flow as classical thermodynamics demands. 
Second, $S^{\rm cg}$ remains strictly non-decreasing in laboratory- and/or thought-experiments, that suddenly reverse all the velocities (strictly, generalized momenta) in the XY system so as to {\em exactly rewind} the prior trajectory. This is paradoxical since for an evolved-and-reversed system, $S^{\rm cg}_{XY}$ takes different values in initial and final states that are precisely the same, directly contradicting the notion that entropy should be a state function. This `echo paradox', named after the spin-echo experiments that inspired it~\cite{teVrugt2021,Ridderbos1998}, is further discussed later. The conventional route to dynamical entropy via coarse-grained cells fails here because it replaces the exactly reversible Hamiltonian dynamics of the XY system with an altered, irreversible dynamics. Crucially, the MDE makes no such replacement and the paradox is avoided.

By instead discarding the information content of $\rho_{Y|X}$, our MDE shows a second law under normal conditions. Here the X dynamics are insensitive to bath details, while the correlations encoded in $\rho_{Y|X}$ become ever more complicated over time. However, atypical states exist for which the reverse is true, XY correlations unravel, and $S^{\rm md}$ decreases. We are currently unaware of any other microscopic, dynamical definition of entropy that {\em quantitatively} reproduces the second law under normal conditions, {\em and yet} deals with exceptions such as those posed by spin-echo.
In these aspects, the MDE provides a quantitative microscopic explanation of why macroscopic thermal entropy takes its familiar form, and behaves as it does under both normal and echo-like conditions. 

Our MDE allows a quantitative study of the conditions for normal, second-law compliant behaviour and of how this is approached as the system size becomes large. We can be precise about this whenever the  exact Hamiltonian dynamics for the total system XY is \emph{mixing}, a condition slightly stronger than ergodicity~\cite{Walters1982,Cornfeld1994}.  
(Such a restriction is natural, as not every Hamiltonian system exhibits thermal behavior.)
Under mixing, the long-time visiting frequency of X coincides with its equilibrium distribution, which is directly determined by the CPV.  
In this equilibrium case, the proposed MDE, evaluated for the equilibrium distribution $\rho^{\rm eq}_X$ of the X subsystem, recovers the Boltzmann entropy of the full XY system. 

The mixing assumption is sufficient to explain how irreversibility emerges from the exact Hamiltonian dynamics of a large or small system X connected to a bath Y (that can also be large or small). However, mixing alone does not enforce monotonicity of the MDE ({\em i.e.}, a strict second law): rather, the MDE increases nonmonotonically during relaxation and fluctuates around a plateau value after equilibration.
This behavior of mixing systems is fully consistent with both classical and stochastic thermodynamics arguments~\cite{Seifert2012} and already resolves the echo paradox~\cite{teVrugt2021,Ridderbos1998}, showing that microscopic reversibility does not preclude macroscopic irreversibility.

A qualitatively stronger form of irreversibility emerges when time-scale separation is present alongside mixing.  
If the variables in X evolve on a much slower time scale than the unmonitored degrees of freedom in Y, the reduced dynamics of X becomes effectively Markovian.  
In this regime, fast relaxation processes are, in effect, coarse-grained away by the dynamics itself, and the MDE increases monotonically in time (a strict second law),  in accordance with macroscopic expectations.  
Physically, while the MDE still fluctuates during microscopic relaxation, time-scale separation suppresses these fast fluctuations, so that entropy production becomes strictly non-negative on the slow time-scales of observation.

Time-scale separation, albeit not always perfect, is ubiquitous among Hamiltonian systems with many degrees of freedom, where
macroscopic (emergent or collective) observables  evolve much more slowly than the microscopic variables.  We will explore a particularly transparent illustration, which allows quantitative numerical study of the role of system size in the approach to thermodynamic behaviour. This is the `box-diffusion problem' in which X is the centre-of-mass coordinate of $N$ particles obeying Hamiltonian dynamics in a confining, thermally isolated, box. We expect strict time-scale separation only at large $N$ but find that, even in its absence ($N\simeq 10$) the fluctuating MDE trends upwards during relaxation and plateaus at equilibrium, quantifying the emergence of quasi-thermal behaviour even in a small system.

Our MDE framework connects Hamiltonian dynamics to thermodynamics via mixing and time-scale separation, without any stronger assumptions of emergent stochastic dynamics.  In the present paper we focus on systems evolving on a fixed energy shell, so that the total energy $E = H_{XY}$ of the combined XY system is conserved. Within this setting we show how the MDE accounts for irreversible relaxation and clarifies the microscopic origins of thermal entropy and heat flow. Constancy of $E$ precludes work interactions which are also central to classical thermodynamics. To address these requires a more general formulation allowing a time-dependent Hamiltonian $H_{XY}(t)$ and probability distributions that span multiple energy shells. This extended formalism will be developed in a companion paper~\cite{ding2026MDE2}, hereafter referred to as Paper II. There we also show how the MDE, for systems undergoing time-dependent work interactions, reproduces the key fluctuation theorems of stochastic thermodynamics.

The rest of this Paper is organized as follows. Section~II gives the basic definitions of the MDE, clarifies its relations with other entropies, and explains the physics behind it. In Section~III we study the dynamical properties of the MDE. We show that the CPV gives the equilibrium distribution and that the MDE increases monotonically under time-scale separation. In Section~IV we illustrate the MDE framework using examples. Section~V concludes with a discussion of implications and future directions, including the possibility of generalizing our definition of the MDE to the quantum domain.

\section{Microscopic Dynamical Entropy: Definition and physical meaning}
\label{sec:MDE-2nd}
Consider the coupled Hamiltonian dynamics of a system comprising X and Y with Hamiltonian $H_{XY}(\xv,\yv)$. Its dynamics is described by the joint probability distribution $\rho_{XY}(\xv,\yv,t)$, where $\xv$ and $\yv$ include both the positions and momenta of the X and Y subsystems respectively. 
Setting $k_B=1$, the Gibbs entropy of the full system,
\ba
S^G_{XY}(t) = - \int \rho_{XY}(\xv,\yv,t)  \log \rho_{XY}(\xv,\yv,t)  d\xv  d\yv,
\label{gibbs-0}
\ea
is constant in time by Liouville's theorem~\cite{Landau1976,Sethna2021}. 

We define the Microscopic Dynamical Entropy (MDE) as
\ba
S^{\rm md}_{XY}(E,t) &\equiv& S^G_X(t) + \int \rho_X(\xv,t) \log \Omega_Y(\xv,E) d\xv  \label{def-mde-minus}
\\& = & - \int \rho_X(\xv,t) \log \frac{ \rho_X(\xv,t)}{ \Omega_Y(\xv,E)} d\xv,
\label{def-mde-0}
\ea
where $S^G_X = - \int \rho_X \log \rho_X  d\xv$ is the Gibbs entropy of subsystem X, with $\rho_X(\xv,t) = \int \rho_{XY}d\yv$ the marginal distribution. The second term in \eqref{def-mde-minus} is the X-averaged Boltzmann entropy of the Conditional Phase-space Volume (CPV) of subsystem Y, which we defined as
\ba
\Omega_Y(\xv , E) &\equiv& \int d\yv  \delta (E - H_{XY}(\xv , \yv)).
\label{Omega-y-def-0}
\ea
Note that the MDE is {\em not} a Kullback-Leibler divergence since $\Omega_Y$ is not a normalized distribution. 

The parameter $E$ in \eqref{Omega-y-def-0} is the total energy of the XY system. For simplicity in this Paper we assume the total Hamiltonian $H_{XY}$ is time-independent, so the XY system can be assigned a definite and constant $E$ value; for generalizations to time-dependent protocols see Ref.~\cite{ding2026MDE2}.

Note that in \eqref{def-mde-0} the roles of X and Y are asymmetric: typically we regard Y as a background or bath system, and are then interested in the marginal distribution $\rho_X$. 
In thermodynamics, the internal microscopic details of Y are unobserved and irrelevant. Only its temperature and total energy are important, as reflected in the expression of the thermal entropy increment as $- \Delta Q/T$. 
This discarding of information about the bath lies at the fundamental origin of irreversibility.
Nonetheless, the definition \eqref{def-mde-0} of $S^{\rm md}$ is completely general, in the sense that for an arbitrary Hamiltonian system with two or more degrees of freedom, we can always partition these into two sets X and Y. Thus the MDE formalism is applicable to even a small isolated Hamiltonian system, as long as we identify as X a specified subset of its degrees of freedom (usually slow variables). This will be illustrated, using the box diffusion problem as an example, in Sec.~\ref{sec:examples}.

Our definition of the MDE depends on the system-bath partition; this aligns with physics as described by thermodynamics. Thermodynamic entropy is not {\em a priori} an intrinsic property of a microscopic state, but quantifies the amount of work that can be extracted under specified controls~\cite{Pippard1966} (see also ~\cite{ding2026MDE2}). Consequently, any microscopic definition of entropy should depend on which degrees of freedom are treated as system variables ($\xv\in$ X), and which are regarded as unobserved, uncontrolled, or environmental ($\yv\in$ Y). This dependence is usually left implicit in thermodynamics, and also in many stochastic descriptions, where the system-bath partition is fixed from the outset. 

In the present framework, the subdivision into X and Y is not itself unique or fundamental. Rather, it encodes an operational distinction between variables that are monitored or manipulated and those that are not. The MDE makes this dependence explicit: different choices of X correspond to different thermodynamic descriptions, and the  MDE quantifies irreversibility, and heat flow, subject to that choice. In this sense, irreversibility is not attributed to a breakdown of microscopic reversibility, but emerges from the structure of the system-bath partition that defines the relevant degrees of freedom. The only structural requirement is that the full set of variables $(\xv,\yv) \in$ XY defines a phase-space representation in which the measure is preserved under the dynamics. In particular, this condition is satisfied when the dynamics is Hamiltonian and $(\xv,\yv)$ are canonical coordinates. More generally, however, the choice of X can be arbitrary, provided that the definition of the CPV, $\Omega_Y$, is generalized accordingly.
For simplicity, we postpone this generalization until Sec.~\ref{sec:general-x}. Until then, we assume that for the variables $\xv$ of interest, one can always construct corresponding variables $\yv$ such that $(\xv,\yv)$ forms a measure-preserving representation.

\subsection{Connection with Gibbs entropy}
\label{sec:t-depend}
To see the physics behind the MDE more clearly, we now derive the relationship between the conserved Gibbs entropy $S^G_{XY}$ and the MDE. 
We decompose the joint distribution as
\ba
\rho_{XY}(\xv,\yv,t) = \rho_X(\xv,t)  \rho_{Y|X}(\xv,\yv,t),
\ea
where $\rho_{Y|X}$ is the conditional distribution of $\yv$ given $\xv$. 
The motivation is that, if a time-scale separation exists between a fast bath Y and a slow subsystem X (which we will discuss in detail in Sec.~\ref{sec:2nd-law}), then on intermediate time scales the marginal $\rho_X$ changes little while $\yv$ fluctuates rapidly. The conditional distribution $\rho_{Y|X}$ thus characterizes the fast bath dynamics.

The evolution of $\rho_{Y|X}$ is essential for the conservation of the total Gibbs entropy. 
Since
\ba
S^G_{XY} & = & S^G_X + S^G_{Y|X}, \\
S^G_{Y|X}& = &- \int \rho_X \rho_{Y|X}  \log \rho_{Y|X}  d\xv  d\yv,
\ea
the difference between the Gibbs entropy and the MDE corresponds to replacing $S_{Y|X}$ by the average log-CPV.
But it is precisely the term $S^G_{Y|X}$ that, by compensating changes in $S^G_X$, keeps $S^G_{XY}$ constant in time. 
In contrast, $S^{\rm md}_{XY}$ depends only on $\rho_X$ and contains no information about $\rho_{Y|X}$, beyond that implied by the conservation of the total energy $H_{XY} = E$.

 \begin{figure}
\includegraphics[width=\columnwidth]{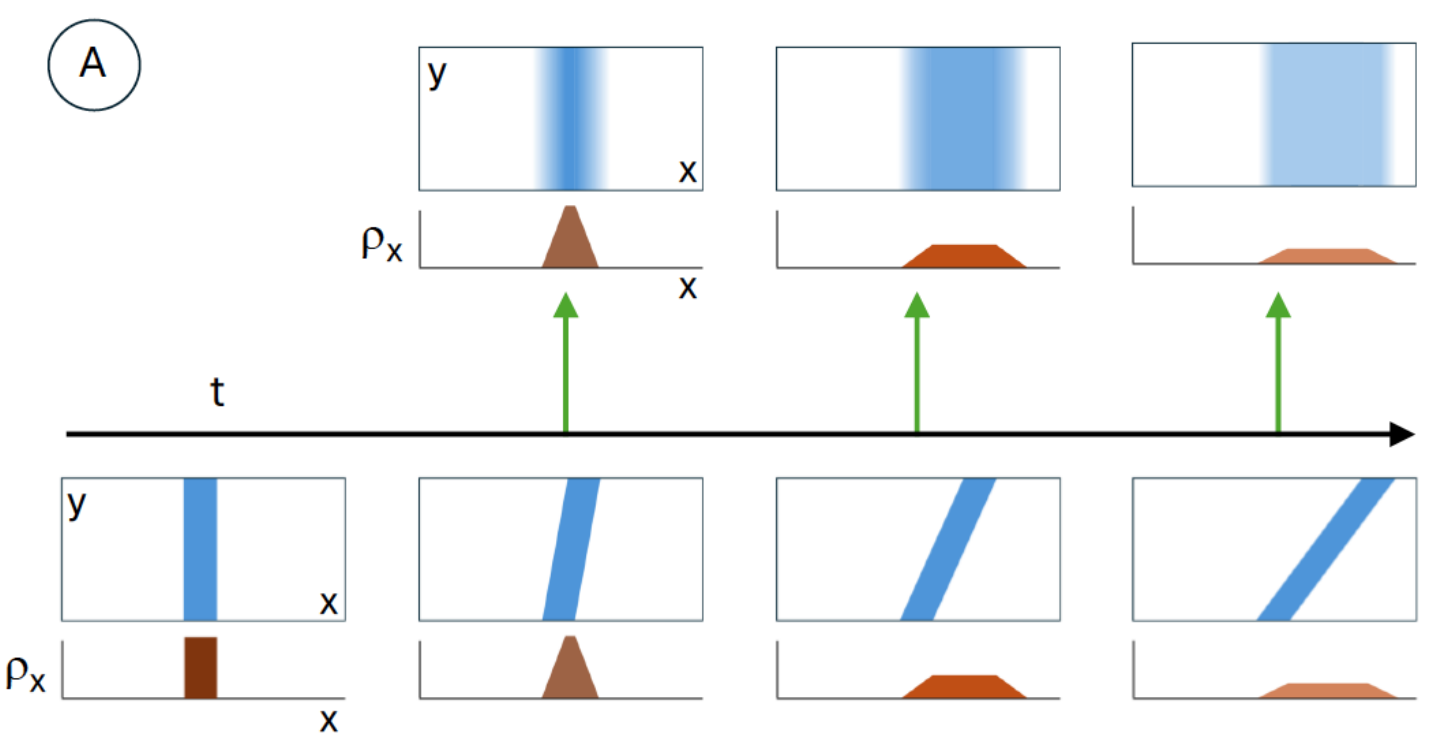}
$$ $$
\includegraphics[width=\columnwidth]{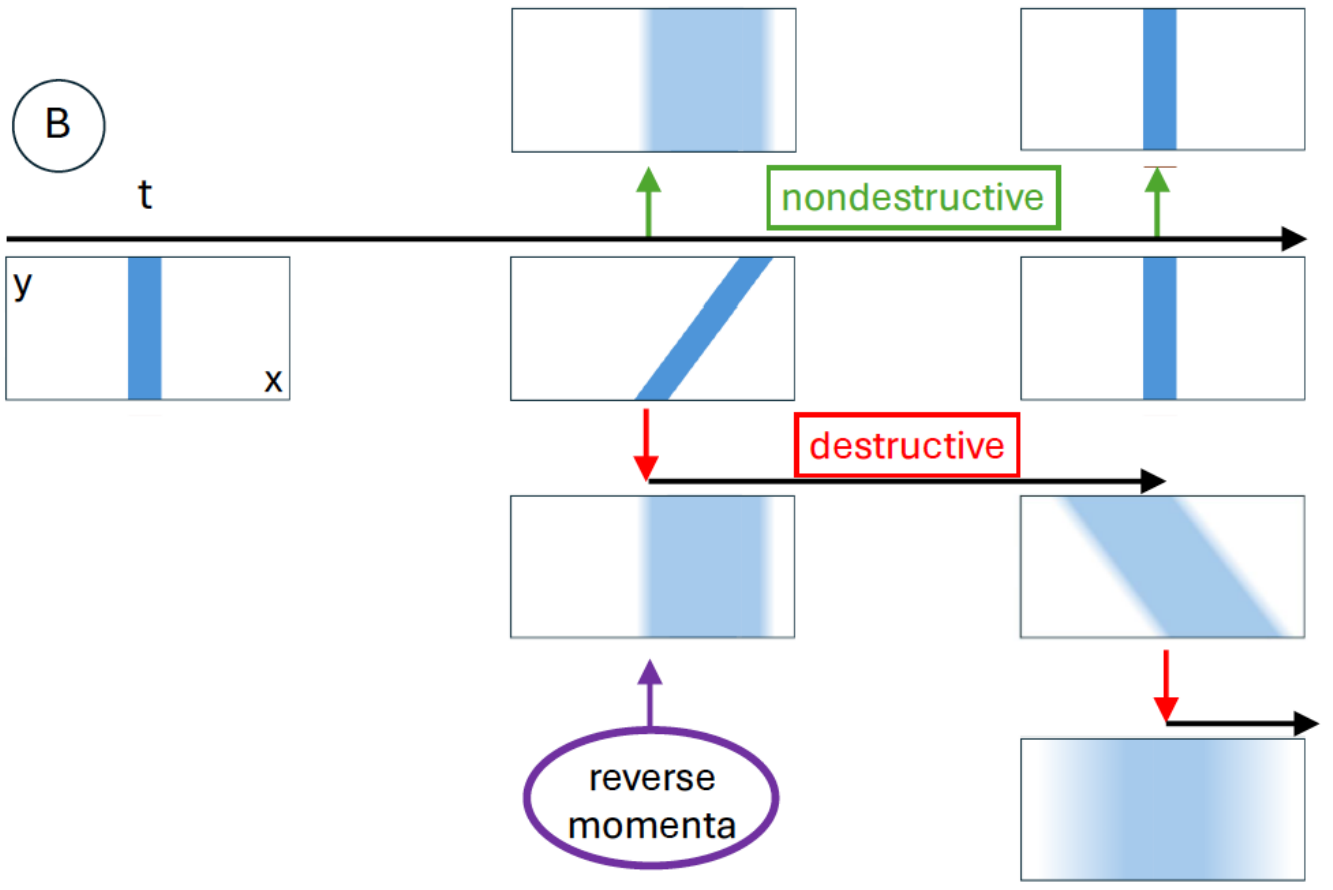}
\par
\caption{Schematic evolution of full and coarse-grained densities in phase space. 
{\bf (A)} Evolution of the full density $\rho_{XY}$  (blue) and marginal $\rho_X$ (brown) in the phase space of an XY system. In this simple example, time evolution shears the distribution under Hamiltonian dynamics, preserving total entropy but increasing the correlations encoded in $\rho_{Y|X}$. Our nondestructive coarse-graining (green arrows) homogenizes $\rho_{Y|X}$ without altering $\rho_X$ or the dynamics.
{\bf (B)} Echo protocol. The middle column shows the same final state as in (A); velocity reversal then restores the initial state under exact Hamiltonian dynamics. Nondestructive coarse-graining (green) preserves the echo; destructive coarse-graining (red), as used in Gibbsian phase-cell methods~\cite{Ehrenfest1959,Gibbs1902,Maes2003}, breaks reversibility and suppresses the echo.}
\label{fig1}
\end{figure}

We now show that $S^{\rm md}_{XY}$ equals to the Gibbs entropy $S^G_{XY}[\tilde\rho]$ of a non-destructively coarse-grained distribution: see Fig.~\ref{fig1}A. The distribution in question is
\begin{subequations}
\ba
&&\tilde \rho_{XY} = \rho_X  \tilde \rho_{Y|X}(\xv,\yv), \\
&& \tilde \rho_{Y|X}(\xv,\yv)  =  \frac{\delta ( E - H_{XY} ) } { \Omega_Y(\xv , E) } .
\ea
\label{cg-00}
\end{subequations}
The coarse-graining homogenizes over Y the conditional probability $\rho_{Y|X}$. It is used only to calculate the MDE and not, as in phase-cell methods, to alter dynamical outcomes.
We can calculate the Gibbs entropy of this coarse-grained distribution as
\ba
&&S^G_{XY}[\tilde \rho_{XY}]   \nonumber\\
&=& - \int \rho_X \tilde\rho_{Y|X}  \log(\rho_X\tilde\rho_{Y|X})  d{\xv} d{\yv} \nonumber\\
& = & S^G_X - \int  \rho_X  d\xv \int \frac{ \delta ( E - H_{XY} ) }{ \Omega_Y(\xv , E) }
 \log  \frac{ \delta ( E - H_{XY} )}{ \Omega_Y(\xv , E) }  d\yv \nonumber\\
& = & S^G_X + \int  \rho_X \log   \Omega_Y(\xv , E)  d\xv + C,
\label{cg-entropy}
\ea
where $C$ is an (infinite) additive constant independent of $E$. 
The latter can be regularized by replacing the $\delta$-function in Eq.~(\ref{cg-00}) with a narrow Gaussian of width $\epsilon$, giving $C = -\log \epsilon$. 
Such a constant is a standard feature of microcanonical entropies ({\em e.g.},~\cite{Sethna2021}), is physically irrelevant, and will be omitted in what follows. As will become clear in Paper II, this contribution has a well-defined origin in terms of a more general energy distribution $\rho(E)$. Its divergence here is a consequence of restricting the dynamics to a single energy shell. In the more general case, the corresponding term remains finite and acquires a direct physical interpretation.

We thus obtain the relation
\ba
S^G_{XY}[\tilde\rho_{XY}] = S^{\rm md}_{XY}[\rho_{XY}] .
\label{g-mde}
\ea
The physical meaning of this relation is transparent: we disregard the detailed information encoded in $\rho_{Y|X}$ for a given $\xv$, by replacing this with $\tilde \rho_{Y|X}$ defined in \eqref{cg-00}, which is uniform in $\yv$. This reflects the thermodynamic viewpoint that thermal entropy ignores the microscopic details of the reservoir, as encoded in the conditional probability of its coordinates, $\rho_{Y|X}(\yv)$. In this sense, the equality in Eq.~(\ref{g-mde}) can be optionally regarded as an alternative and fully equivalent definition of our MDE.

Crucially, in this approach to coarse-graining, the underlying Hamiltonian dynamics remains unchanged: the marginal distribution $\rho_X$ still evolves according to the exact Hamiltonian dynamics of the full XY system. Thus interpretation \eqref{g-mde} represents a {\em non-destructive} coarse-graining, as defined in the Introduction and shown in Fig.~\ref{fig1}. As illustrated there and discussed further in Section~III, this means that the second law runs backwards after velocity reversal, as it must do if entropy is a state function. Crucially also,  Eq.~(\ref{g-mde}) does not mean that we assume the bath to be actually governed by  the flat conditional distribution $\tilde \rho_{Y|X}$. On the contrary, in `thermodynamically typical' systems, the bath is governed by a time-dependent conditional distribution $\rho_{Y|X}(t)$ of ever-increasing complexity. It is precisely by replacing this (in the entropy calculation but {\em not} the dynamics) with the flattened form $\tilde \rho_{Y|X}$, that the MDE can increase even as $S^G_{XY}$ stays constant. In Sec.~\ref{sec:2nd-law} we will show this increase to be monotonic, under assumptions that are reasonable for thermodynamic systems.

\subsection{Connection with the thermal entropy}
\label{sec:firstlaw_traj}

In what follows, we show how the MDE fulfils the defining properties of the conventional thermal entropy. For clarity, we first consider the weak-coupling regime, corresponding to the standard setting of textbook thermodynamics where the interaction energy between the observed variables X and the unobserved variables Y can be neglected in the calculation of the conditional phase volume.

This weak-coupling assumption is introduced here only to facilitate comparison with conventional thermodynamics. The general case will be treated in Paper II~\cite{ding2026MDE2}, where weak coupling, large-bath approximations, or any other additional assumptions, are not invoked.

We write the total Hamiltonian in the form
\ba
H_{XY}(\xv,\yv)
=
H_X^0(\xv)
+
H_Y^0(\yv)
+
H_I^0(\xv,\yv).
\label{three_terms}
\ea
By weak coupling we mean that, in the calculation of the CPV, the interaction term $H_I^0$ does not contribute at leading order. In this case,
\ba
\Omega_Y(\xv,E)
=
\int d\yv\,\delta(E-H_{XY})\nonumber\\
\approx
\int d\yv\,\delta(E-H_X^0-H_Y^0).
\ea
We define the bath energy as
\ba
E_Y(\xv)
\equiv
E-H_X^0(\xv),
\ea
so that the CPV becomes a function of $E_Y$ only. Thus we define
\ba
\Omega_Y(E_Y)
&=&
\int d\yv\,\delta(E_Y-H_Y^0(\yv)),\\
S_Y^\Omega(E_Y)
&\equiv&\log\Omega_Y(E_Y),
\ea
and the corresponding inverse temperature
\ba
\frac{1}{T}
=\beta
\equiv
\frac{\partial S_Y^\Omega}{\partial E_Y}.
\label{bath_temp}
\ea
In the large-bath limit where $E_Y \gg H_X^0$, the MDE then simplifies to
\ba
S^{\rm md}_{XY}
=S_X^G
+ \langle S_Y^\Omega\rangle_X
= S_X^G
+ \langle \beta E_Y\rangle
+ {\rm const.}
\label{NTE-0}
\ea
or equivalently
\ba
S^{\rm md}_{XY}
= S_X^G
- \langle \beta H_X^0\rangle
+ {\rm const.}.
\label{NTE-1}
\ea
In classical thermodynamics, apart from the additive constant $\langle S_Y^\Omega(E)\rangle_X$, the expression in (\ref{NTE-1}) coincides with $-A_X/T$, where the availability is conventionally defined as $A_X \equiv E_X-TS_X$ for a system at fixed volume in contact with a bath at temperature $T$~\cite{Pippard1966}. Indeed, $-A_X/T$ represents the total thermodynamic entropy of the combined system in this setting, and the second law implies $\dot A_X\le0$. Near equilibrium, where a well-defined temperature can also be assigned to X itself, $A_X$ reduces to the Helmholtz free energy $F_X=E_X-TS_X$~\cite{Pippard1966}.

\subsection{Extension beyond fixed energy}
Throughout this paper, we restrict attention to autonomous systems on a fixed energy shell. This allows us to focus on the statistical origin of irreversibility while avoiding additional complications associated with external driving. As a result, the total energy remains constant and does not yet play an explicit dynamical role. The framework developed here, however, admits a natural extension beyond the fixed-energy setting. In Paper II, the total energy $E$ is promoted to an explicit state variable, leading to a generalized entropy functional 
\ba 
S^{\rm md}_{XY}(t) &=& S_E^G[\rho_E] + \int dE\, \rho_E(E,t)\log Z(E,t) 
\label{mde-kl-form}\\ 
&-& \int dE\, \rho_E(E,t) D\left( \rho_{X|E}(\xv,E,t) \middle\| \rho_X^{\rm ref}(\xv,E,t) \right). \nonumber 
\ea 

The significance of this extension is not merely the appearance of a more general entropy formula. Rather, it identifies $(X,E)$, instead of X alone, as the natural choice of thermodynamic state. Since the total energy automatically records energy exchange with external driving, work interactions are incorporated at the microscopic level without introducing any additional thermodynamic assumptions. This viewpoint remains well defined for arbitrary Hamiltonian systems, including strongly coupled systems and finite reservoirs. In particular, the resulting theory yields exact fluctuation relations even for systems consisting of only a few degrees of freedom. The only additional physical assumption required for the emergence of irreversible thermodynamic behavior is a separation of time scales between the observed and unobserved variables. The present paper establishes the microscopic entropy structure underlying this program. Its thermodynamic consequences, including heat, work, fluctuation relations, and the emergence of the second law, will be developed in Paper II.

\vspace{5mm}
\section{Time-evolution of MDE and Hamiltonian irreversibility}
\label{mecsec:irrev}

In this Section we consider the dynamical properties of the MDE. First we show how the conditional phase-space volume, $\Omega_Y(\xv)$, is related to the equilibrium distribution. Then we demonstrate that under assumptions that prevail in thermodynamics, the MDE increases monotonically until equilibrium is reached. Finally we briefly talk about how MDE emerges dynamically under time-scale separation. As already emphasized, the MDE, which is a state functional of $\rho_X$, retains Hamiltonian reversibility while explaining the second law at the same time.

\subsection{Relaxation to equilibrium}
\label{sec:mixing}
To understand the physical meaning of $\Omega_Y$, consider the case where total Hamiltonian $H_{XY}$ is time-independent and the dynamics of XY is mixing on the relevant energy shell~\cite{Walters1982,Cornfeld1994}. This means that any initial patch of phase space is stretched and folded until its image becomes uniformly distributed with respect to the Liouville measure. In contrast to the phase-space distribution of the total system $\rho_{XY}$ whose Gibbs entropy is conserved, the marginal distribution $\rho_X$ relaxes to 
\ba
\rho_X^{\rm eq}(\xv)= \frac{ \Omega_Y(\xv)}{\int \Omega_Y(\xv) d\xv }. \label{mec_micro}
\ea
Here the normalization constant is set by the Boltzmann entropy $S^B_{XY}$ of the total system:
\ba
\int \Omega_Y(\xv)\, d\xv = \int \delta(E - H_{XY})d\xv d\yv = \exp\left[S^B_{XY}\right].
\ea
For the MDE at equilibrium we then find from (\ref{def-mde-0}, \ref{mec_micro})
\ba
&& S^{\rm md}_{XY} [\rho_X^{\rm eq}(\xv) ] = S^B_{XY}. \label{mec_relax}
\ea

The CPV $\Omega_Y(\xv)$ can be viewed a geometric quantity determined by the total Hamiltonian $H_{XY}$. It gives different points $\xv$ different statistical weights. Under the assumption of mixing, these coincide with the temporal weight with which each point $\xv$ is visited in the long time limit. Thus in a finite, mixing, Hamiltonian system the relaxation towards $\rho_{XY}^{\rm eq}$ is established without approximation and therefore so is the relaxation of the MDE towards its equilibrium value obeying \eqref{mec_relax}. Moreover, since $S^{\rm md}_{XY}$ is found from $S^G_{XY}$ by homogenizing the conditional density $\rho_{X|Y}$, it is bounded above by $S^B_{XY}$ and therefore approaches that value from below.
(Note that for it to approach monotonically is a stronger condition, addressed in Sec.~\ref{sec:2nd-law} below.) In this sense the MDE already captures the emergence of irreversibility and relaxation in Hamiltonian dynamics, so long as the XY system is mixing.

Notably, the relaxation of $\rho_X$ to $\rho_X^{\rm eq}$ --- and hence the relaxation of $S^{\rm md}_{XY}$ to $S^B_{XY}$ ---  does not contradict Poincar\'e recurrence. The irreversibility encoded by the MDE concerns the fraction of time a typical trajectory spends in each region, which relaxes to the Liouville measure on the energy shell. Poincar\'e recurrence only asserts that each trajectory eventually returns arbitrarily close to its initial point, but such events have vanishing temporal weight and do not prevent relaxation. In contrast, under mixing dynamics, the empirical distribution --- and equivalently the expectation value of any observable --- relaxes to its equilibrium value determined by $\rho_X^{\rm eq}(\xv)\propto \Omega_Y(\xv)$. (This relaxation is governed by the decay of dynamical correlations, and occurs on a physically sensible timescale, distinct from the unphysically long recurrence or ergodicity time.)

\subsection{Echo paradox resolved}
\label{sec:echo}

The MDE, while it can describe relaxation as just explained, also encodes Hamiltonian reversibility as required by the classical spin-echo experiment~\cite{Ridderbos1998} and related (thought-) experiments (see Fig.~\ref{fig1}B). This is because it is a state function of $\rho_X(\xv)$, which means that any dynamical manipulation of the XY system which restores  $\rho_X$ to an earlier distribution likewise restores the MDE to its earlier value. The `echo paradox', which is a version of Loschmidt's paradox~\cite{Wu1975}, views the resulting entropy decrease as paradoxical, because it violates the second law.    
It is linked to a wider set of `irreversibility problems', informatively surveyed in~\cite{teVrugt2021}. We argue below that the second law {\em should} be violated in echo protocols, and show that the MDE does allow such violations.

A Hamiltonian thought experiment of `echo' type is the removal of a partition at $t = 0$ in an isolated box of gas. The thermodynamic entropy increases as the gas expands. If all particle velocities are reversed at $t = t_r$, then at $t = 2t_r$ the system returns exactly to its initial microstate and the partition can be reinserted with zero work. Since this holds for each microstate it also holds for the phase-space density $\rho_{XY}$, which is likewise restored to its initial condition so that the final state is again the equilibrium state in the left half. (Note that it is not necessary to explicitly identify the X and Y subsystems to make this argument.)

The entropy increase following expansion is therefore unwound by the echo protocol, giving a formal violation of the second law. Such a violation is inevitable if the entropy is to remain a state function (since initial and final states are identical). Therefore, any microscopic interpretation of entropy that insists on a monotonic entropy increase \emph{even after} velocity-reversal contradicts thermodynamics rather than explaining it. Gibbsian phase-cell coarse-graining dramatically fails this test. In contrast, our MDE correctly rewinds under reversal  of all velocities in the XY system (Fig.~\ref{fig1}). 

The phase-cell approach fails in echo protocols because its coarse-graining alters the dynamics of $\rho$ (Fig.~\ref{fig1}B). Even if the chosen cell-size reflects a genuine resolution limit of practical observations, its destructive coarse-graining procedure contradicts the precept that in classical mechanics, observations -- whether vague or precise -- do not alter trajectories. Similar objections apply to other dynamical approximations involving projection operators, dynamic density functionals, Fokker-Planck equations, {\em etc.}~\cite{Zwanzig2001,teVrugt2021}. These also do not describe the dynamics after velocity-reversal, because they approximate the reversible,  underlying Hamiltonian dynamics with an explicitly irreversible time evolution.

The echo paradox applies to all isolated systems, including those of XY type studied here. For these systems in particular, it is resolved by our MDE. Importantly also, we establish below conditions under which the second law (monotone increase of entropy with time, Eq.~\eqref{SecondL} below) emerges for the MDE. These conditions require that the influence of Y on X is  `bath-like', so that the X dynamics is autonomous in a sense we shall define. Subsystem Y clearly ceases to be bath-like upon velocity-reversal: instead it meticulously unravels its correlations with X. Indeed, replacing the bath with a newly prepared one of equal temperature would (unless the XY coupling is truly negligible) generically erase the echo.

\subsection{Autonomy and monotonicity of the MDE}
\label{sec:2nd-law}

When the heat bath Y is large but finite, we can expect the MDE to be non-decreasing most of the time but to show downward excursions that become exceedingly rare for a sufficiently large system and/or time-scale separation~\cite{Davies1994,Linden2009,Strasberg2017,Pavliotis2008}. To derive a strict second law from the MDE, these excursions must be discarded (effectively sending the time interval between them to infinity) by introducing an additional physical assumption. Clearly that assumption {\em should be violated} by echo experiments, but should prevail in regimes of `normal' thermodynamic behaviour.

To achieve such a derivation,  we now assume that the influence of Y on $\rho_X$ is truly `bath-like': it is conveyed by a sub-extensive set of parameters (such as $T$ and some friction coefficients) rather than by the precise state of Y. More formally, we assume the evolution $\rho_X$ is autonomous in the following sense: if two distributions $\rho_{XY}^{(1)}$ and $\rho_{XY}^{(2)}$ share the marginal
$
\rho^{(1)}_{X}(t_0) = \int   \rho^{(1)}_{XY}(t_0)d\yv
=  \int  \rho^{(2)}_{XY}(t_0)d\yv = \rho^{(2)}_{X}(t_0) 
$
at initial time $t_0$, then for all $t>t_0$:
\ba
\rho^{(1)}_X(t) = \rho^{(2)}_X(t) + O(\epsilon)\simeq \rho^{(2)}_X(t) . 
\label{bath-req-2}
\ea
Here $\epsilon$ is a small parameter, in some cases precisely identifiable via a time-scale separation between X and Y~\cite{Ding2023,Jarzynski1995}, and connected inversely with the excursion time mentioned above.

An example of when the autonomy condition \eqref{bath-req-2} holds true is where $\rho_X$ satisfies a Fokker-Planck equation. 
This outcome can be derived from the Hamiltonian dynamics through the standard multi-scale expansion~\cite{Zwanzig2001,mori1965,Pavliotis2008} under the assumption of time-scale separation, and in this case $\epsilon$ represents the ratio of time scales involved.

To prove the second law from Eq.~(\ref{bath-req-2}), consider an initial density $\rho_{XY}(t_0)$ and its coarse-grained form $\tilde\rho_{XY}(t_0)$ obeying \eqref{cg-00}. By construction these have the same MDE, $S^{\rm md}_{XY} = S^G_{XY}[\tilde\rho_{XY}(t_0)]$. Now evolve $\tilde\rho_{XY}(t_0)$ under $H_{XY}$ to time $t>t_0$. The Gibbs entropy of $\tilde\rho_{XY}$ remains constant as it evolves into $\rho^*_{XY}(t)$; note that this {\em no longer} obeys \eqref{cg-00}. Now coarse-graining a second time  gives
\ba
S^{\rm md}_{XY }[ \rho^*_{XY}(t)]&\equiv& S^G_{XY }[ \tilde \rho^*_{XY}(t)] \nonumber\\
\geq S^G_{XY }[ \rho^*_{XY}(t)] &=& S^{\rm md}[\rho_{XY}(t_0)].\label{coarse}
\ea
The inequality holds because replacing $\rho^*_{Y|X}$ by a uniform distribution (in $\yv$, see \eqref{cg-00}) cannot decrease its entropy.

We now invoke \eqref{bath-req-2} to say that $\rho^*_{XY}(t)$ has the same $\rho_X(t)$ as $\rho_{XY}(t)$ and hence the same $S^{\rm md}_{XY}$. Here $\rho_{XY}(t)$ is the time-evolved full density. The second law
\ba
S^{\rm md}_{XY} [  \rho_{XY}(t)] \geq S^{\rm md}_{XY} [ \rho_{XY}(t_0)]\quad ;\quad t>t_0 \label{SecondL}
\ea
then follows. 
Note that we did not assume equilibrium at the initial and/or final time~\cite{Jaynes1965}. Our only assumption is that of autonomy, as encoded in Eq.~(\ref{bath-req-2}). 

Autonomy does not imply that Y remains at conditional equilibrium. Rather, the deviation of $\rho_{Y|X}$ from equilibrium is governed by $\rho_X$ and is essential for generating the irreversible and stochastic components of the effective dynamics. Indeed, a destructive coarse-graining that sets $\rho_{Y|X}=\rho^{\rm eq}_{Y|X}$ therefore yields only the leading-order drift and, in general, gives an incorrect evolution for $\rho_X$.
This perturbative argument provides an insightful viewpoint on the second law. Eq.~(\ref{g-mde}) equates the MDE to the Gibbs entropy evaluated with an equilibrium bath of the correct energy, whereas the actual dynamics of $\rho_X$ is governed by a bath that deviates slightly from equilibrium. The resulting mismatch breaks the Liouvillian conservation of entropy and allows the MDE to increase.

We stress that this assumption does not require every variable in Y to be fast compared with those in X: the unmonitored variables in Y may include one or more slow variables.  What is required is only that, on the time scale on which the relaxation of X is being described, these variables do not have to be retained in order to obtain a closed autonomous dynamics for X.  This can happen in two distinct ways.  First, a slow variable in Y may evolve on a time scale much longer than the observation time, so that it acts effectively as a fixed parameter or sector label during the relaxation of X.  Second, as in the non-ergodic examples discussed below in Sec.~\ref{sec:information}, a variable in Y may be dynamically decoupled from X, so that its value does not affect the autonomous evolution of the latter.  In either case, the relevant condition is not that all degrees of freedom in Y are fast, but that the variables omitted from X do not introduce memory or additional state-dependence into the effective dynamics of X on the time scale under consideration.

\subsection{Effective dynamics under time-scale separation}
\label{sec:effective-dynamics}

The emergence from Hamiltonian dynamics of an effective stochastic description, for instance at Fokker-Planck level, is a well studied question~\cite{Zwanzig2001,Pavliotis2008,mori1965,Zwanzig1961}. It is conceptually distinct from the question of how to microscopically define a thermodynamic entropy out of equilibrium (the MDE), but the two are closely linked. To explore this we follow here Ref.~\cite{Ding2023} which developed a systematic route from Hamiltonian dynamics to an effective Langevin or Fokker--Planck dynamics for a reduced variable set $\xv$ under time-scale separation. However, that task is not complete until the stationary distribution of the selected variables is correctly identified. 
The present framework supplies this missing thermodynamic input to the approach of Ref.~\cite{Ding2023} and other related approaches.

For a given Hamiltonian system XY and a chosen set of selected variables $(\xv,E)$, the equilibrium structure is determined by the CPV $\Omega_Y(\xv,E)$. For a time-independent Hamiltonian, $E_{XY}=E$ is conserved, and the equilibrium distribution of the selected variables is $\Omega_Y(\xv,E)/Z(E)$. This defines the generalized potential
\ba
U(\xv,E) = -\log \Omega_Y(\xv,E),
\label{U-effective}
\ea
up to an additive constant independent of $\xv$. Unlike phenomenological potentials introduced in stochastic models, $U(\xv,E)$ is determined directly by the phase-space geometry of the underlying Hamiltonian system.

We therefore obtain a complete recipe: the multi-scale projection formalism determines the kinetic part of the effective stochastic dynamics, while the MDE framework determines the stationary distribution and thermodynamic potential through $\Omega_Y(\xv,E)$.

We now briefly recall the dynamical construction from Ref.~\cite{Ding2023}. We assume a separation of time scales between the selected variables $\xv$ and the unobserved variables $\yv$. Let $\epsilon$ denote the ratio between the characteristic time scale of the fast variables and that of the slow variables, with $\epsilon \ll 1$. In this regime, the unobserved variables relax rapidly compared with the evolution of $\xv$, so that over the time scale of $\xv$ they remain close to conditional equilibrium at fixed $(\xv,E)$.

For simplicity, we restrict here to time-independent Hamiltonians $H_{XY}$. (The extension to time-dependent driving will be discussed elsewhere.) Under time-scale separation, the Liouville operator of the full system may be decomposed as
\ba
\mathcal L = \mathcal L_S+\epsilon^{-1}\mathcal L_Y,
\ea
where $\mathcal L_Y$ generates the fast dynamics of the unmonitored $\yv$ variables, while $\mathcal L_S$ contains the slow dynamics and the coupling between $\xv$ and $\yv$.

Applying the multi-scale projection operator formalism of Ref.~\cite{Ding2023}, one obtains a systematic expansion for the reduced dynamics,
\ba
\partial_t \rho_{XE} =
\left(\mathcal L_{XE}^{(0)}+\epsilon \mathcal L_{XE}^{(1)}+\cdots\right)\rho_{XE},
\ea
with
\ba
\mathcal L_{XE}^{(0)} &=& \int d\yv \mathcal L_S e^{-U_Y},
\\
\mathcal L_{XE}^{(1)} &=& \int d\yv \int_0^\infty ds \mathcal L_S e^{s\mathcal L_Y}\mathcal L_S e^{-U_Y}.
\ea
Here $e^{-U_Y}$ denotes the conditional equilibrium distribution of the unobserved variables at fixed $(\xv,E)$. The leading term generates the reversible part of the reduced dynamics, while the first-order correction encodes irreversible effects arising from the relaxation of the unobserved variables. In explicit models, this correction reduces to Green--Kubo-type correlation expressions~\cite{Green1954,Kubo1957,mori1965,Zwanzig1961,Zwanzig2001}.

At leading order in $\epsilon$, the evolution becomes autonomous and Markovian, and takes the Fokker--Planck form
\ba
\partial_t \rho_{XE} =
\partial_i \left[
L_{ij}(\xv,E)
\left(\partial_j+\partial_j U(\xv,E)\right)
\rho_{XE}
\right],
\label{FP-effective}
\ea
where the drift is fixed by the effective potential $U(\xv,E)=-\log \Omega_Y(\xv,E)$, while the kinetic coefficients $L_{ij}$ encode the response of the eliminated variables and can be written, in explicit models, as time-correlation functions of the fast dynamics.

It is useful to compare this result with the Mori--Zwanzig projection formalism~\cite{mori1965,Zwanzig1961,Zwanzig2001}. Mori--Zwanzig gives an exact representation of the reduced dynamics in terms of memory kernels and fluctuating forces. By contrast, the present route assumes an explicit separation of time scales and directly produces a Markovian expansion in the small parameter $\epsilon$. Thus, when the desired effective description is a Langevin or Fokker--Planck equation, the multi-scale construction gives a direct route from the Hamiltonian dynamics, while the MDE supplies the missing equilibrium potential.

Finally, fast relaxation of the unobserved variables does not mean that Y is exactly at conditional equilibrium at all times. The small deviations of the conditional distribution from equilibrium are precisely what generate the stochastic and irreversible terms in the effective dynamics. Time-scale separation therefore justifies an autonomous reduced description, but not a replacement of the full conditional dynamics by equilibrium at every instant as destructive coarse-graining would require. In this setting, the role of the CPV is instead to identify the equilibrium reference state toward which the effective stochastic dynamics relaxes.

\section{Examples: Thermal behavior in isolated small Hamiltonian systems}
\label{sec:examples}
Here we study three different models to show how the MDE quantifies irreversible behaviour in isolated Hamiltonian systems --- even when these are not large. In each case, we implement exact Hamiltonian dynamics: there is no random forcing of any kind. 

We show that thermal behaviour can rapidly emerge as the total particle number $N$ is increased. Relaxation of the MDE towards its equilibrium (maximal) value is seen already at small $N$, with monotonicity of that relaxation approached at larger $N$ when stronger time-scale separation can emerge. This is fully consistent with the discussion given in Section~\ref{sec:2nd-law}. Surprisingly though, we find that near-monotonic thermal relaxation can be seen for $N$ as small as 10 or 20 particles, so long as we choose carefully the partition into X and Y variables. A striking example is when X describes the centre of mass position of particles in a box, as we address next; see Fig.~\eqref{fig:box}.

\subsection{The box diffusion problem}
\label{sec:box}

 \begin{figure}
\includegraphics[width=\columnwidth]{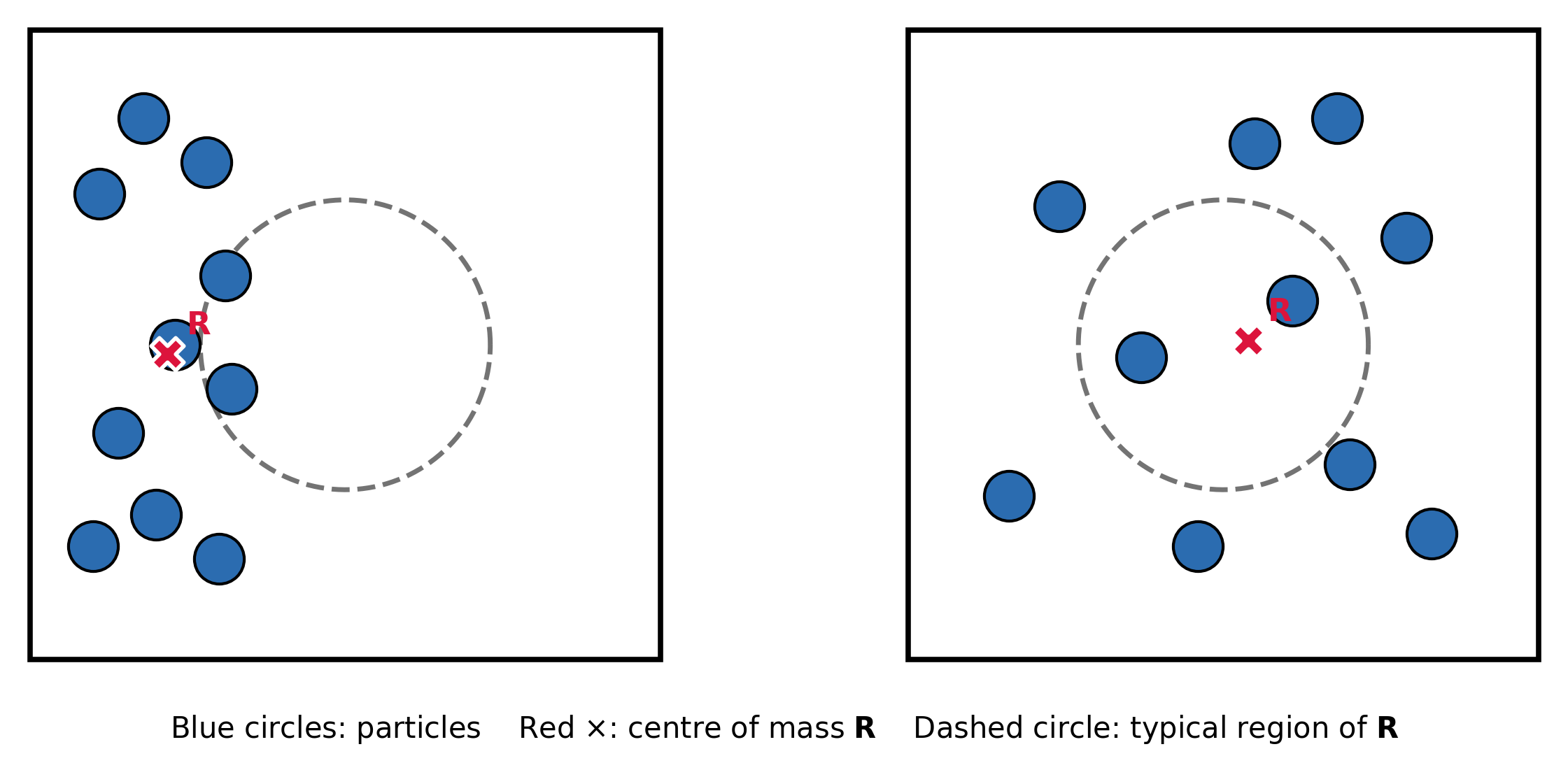}
\par
\caption{Of these two configurations in a gas of particles (colliding elastically with box walls in a closed system), intuition may suggest the right one is a more probable state. However, they are both single points in phase space, and hence equiprobable. What distinguishes them is typicality of collective X-observables such as the centre-of-mass coordinate ${\Rv}$: there are many more configurations with the same $\Rv$ when it is near the center. Such typicality is quantitatively described by $\Omega_Y(\Rv)$. }
\label{fig:box}
\end{figure}

Consider $N$ particles of radius $a$ in a square box $[0,L]^2$, as in Fig.~(\ref{fig:box}).  Two microscopic configurations that differ macroscopically (all particles on one side versus uniformly spread) are both single points in phase space; their difference arises solely from the relative phase-space volume of all microstates compatible with $\xv$. This means that case (b) in Fig.~(\ref{fig:box}) is more typical than case (a)~\cite{Goldstein2006,Popescu2006,Cover1999}, with typicality quantitatively described by $\Omega_Y$.

For definiteness we now choose X to comprise the centre-of-mass (CoM) coordinates of the total system. In other words, we select two X-variables $\Rv \equiv (R_x,R_y)$ where
\ba
\Rv=\frac{1}{N}\sum_{i=1}^N \rv_i,
\ea
and let $\yv \in$ Y include all the relative coordinates $\rv_i-\Rv$ and all momenta $\pv_i$. The Hamiltonian has the form
\ba
H_{XY}
= \sum_{i=1}^N \frac{|\pv_i|^2}{2m}
+ \sum_{i<j} U(|\rv_i-\rv_j|)+ \sum_i V_{\rm wall}(\rv_i).
\ea
We take the interaction $U$ to be (for example) a hard-core repulsion. As usual, the momenta $\pv_i$ can be integrated out independently of $\Rv$, which gives
\ba
\Omega_Y(\Rv,E)=C_N E^{N-1} V^{\rm conf}(\Rv),
\ea
with the configurational part given by
\ba
V^{\rm conf}(\Rv)
&=& \int \Big[\prod_i d^2\rv_i\Big]
\Big[\prod_{i<j}\mathbf 1_{|\rv_i-\rv_j|\ge 2a}\Big] \nonumber\\
&&\delta^{(2)}\!\Big(\tfrac{1}{N}\sum_i \rv_i - \Rv\Big),
\ea
which counts the phase-space-volume of internal configurations consistent with the CoM position $\Rv$. 

The evolution of $\Rv$ is governed by
\ba
\dot \Rv = \frac{ \Pv }{M}
\ea
with $\Pv = \sum_i \pv_i$ the total momentum and $M= Nm$ the total mass. The magnitude of $\Pv$ is constrained by the total energy $E$. For fixed average energy $E/N$ and particle mass $m$, $\dot\Rv$ is typically of order $O(1/\sqrt{N})$, and its motion is slowed by a factor $1/\sqrt N$ compared to the relative coordinates. This timescale separation makes it a good choice for subsystem X. 

\begin{figure}
\includegraphics[width=\columnwidth]{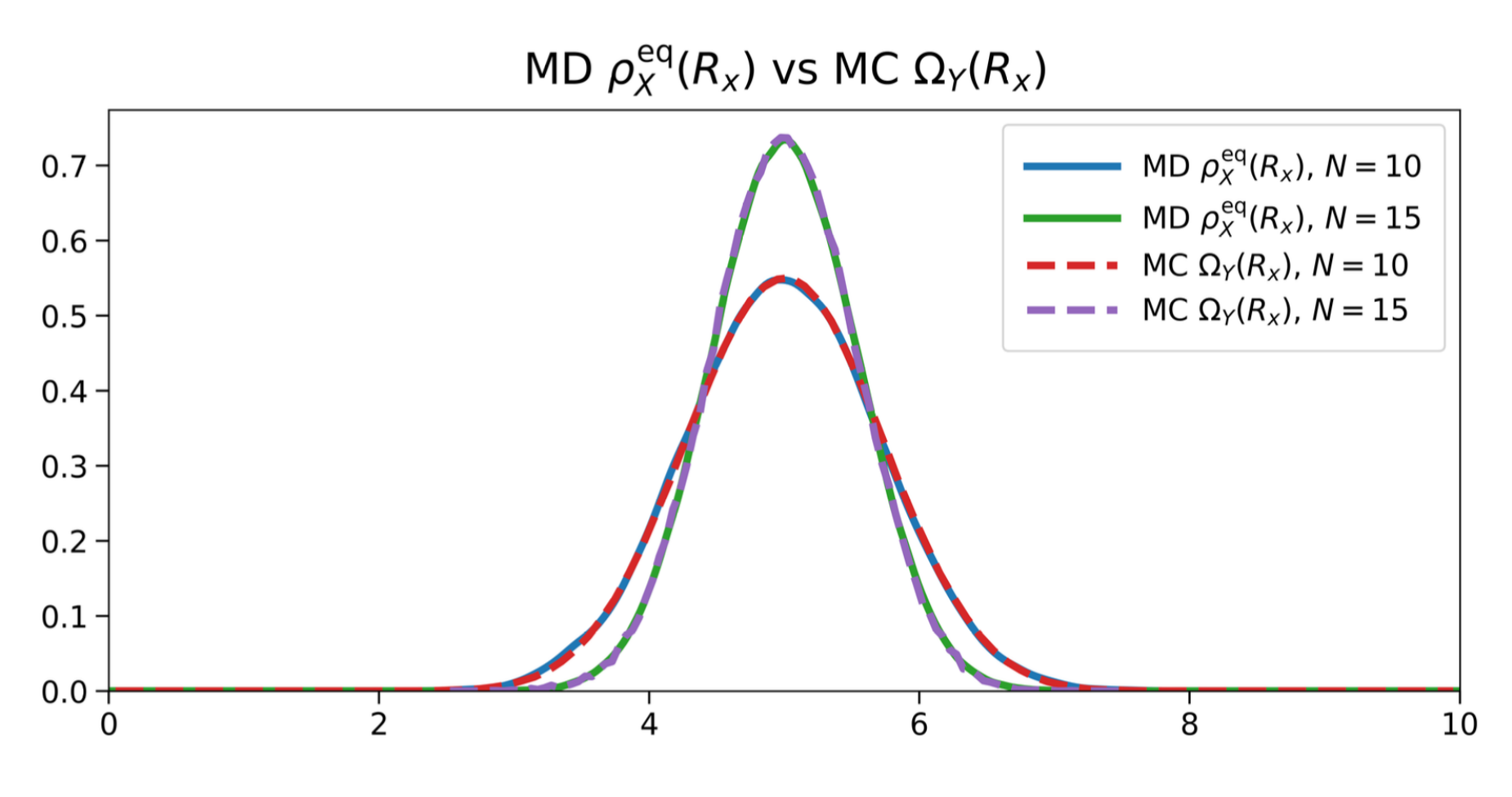}
\par
\caption{For the box diffusion problem, we compare the long time distribution $\rho_{\rm eq}(R_x)$ and the numerical calculation of $\Omega_Y(R_x)$. The latter has been normalized to allow direct comparison. This shows how a fully deterministic Hamiltonian dynamics of the XY system leads to an equilibrium distribution for the slow variable $R_x \in$ X.}
\label{fig:box-10}
\end{figure}

We next compare numerical calculations of $\rho^{\rm eq}_X$ and  $\Omega_Y$. The particle radius is chosen as $a=0.5$ and the size of the box is $L_x = L_y = 10$. The initial velocities are sampled independently from a Gaussian distribution, after which the center-of-mass velocity is subtracted to enforce zero total momentum. The velocities are then uniformly rescaled so that the total kinetic energy is fixed to unity. We use Monto-Carlo integration to calculate $\Omega_Y(\Rv)$ and normalize the result for comparison with $\rho^{\rm eq}_X(\Rv)$. To find the latter quantity, we simulate the deterministic motion using molecular dynamics and sample the statistics of $\Rv$ in the stationary state after a suitable period for the system to relax. We show the equilibrium distribution for $R_x$ in Fig.~\ref{fig:box-10}, with $N=10$ and $15$ separately. The distribution matches accurately the normalized $\Omega_Y$. The curves are Gaussian because the interactions generate only short-range correlations, and $R_x$ obeys the central limit theorem.

\begin{figure}
\includegraphics[width=0.8\columnwidth]{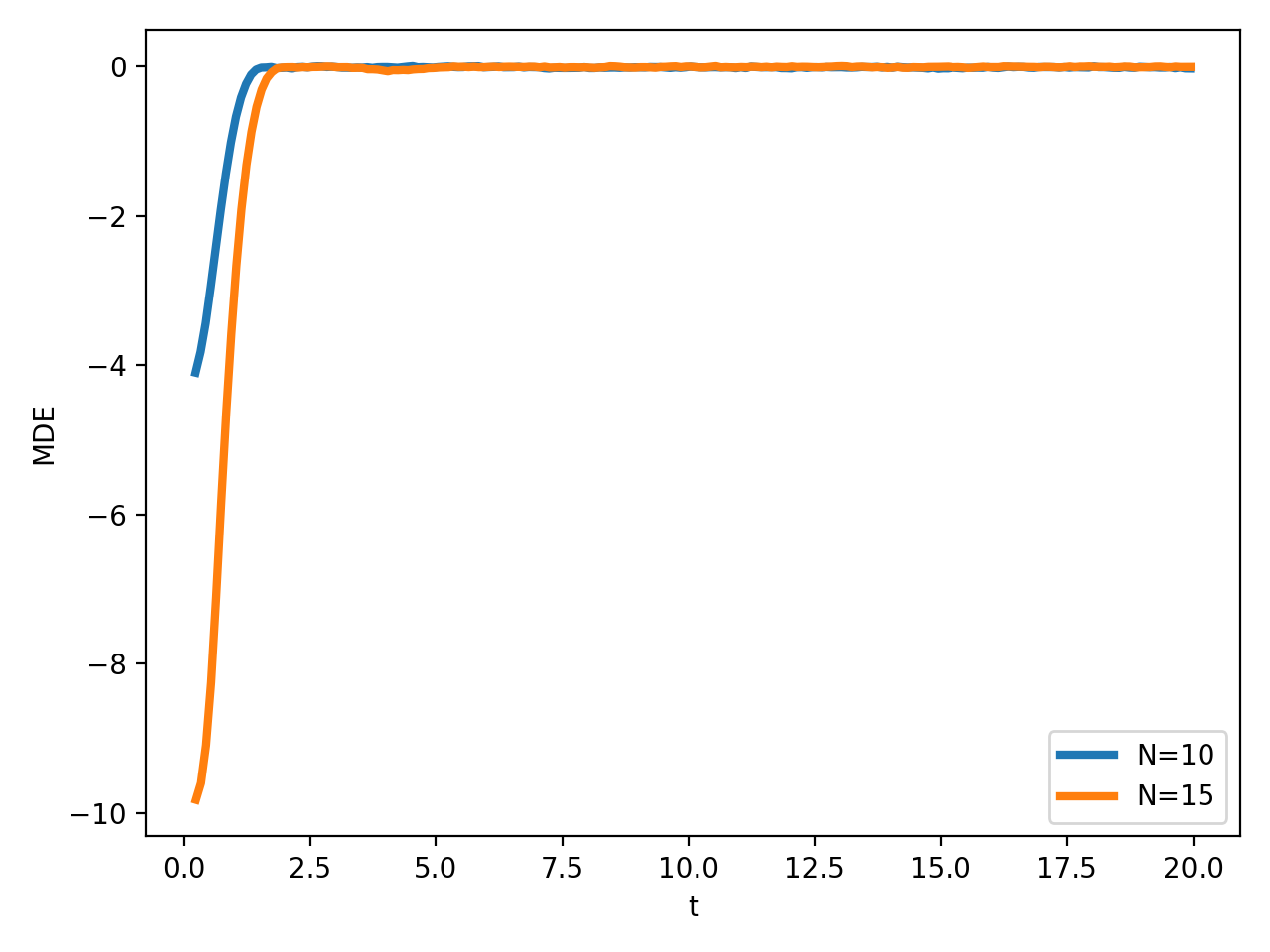}
\par
\caption{Time-evolution of the MDE (measured relative to its equilibrium value) for the box-diffusion problem with $N=10$ and $15$. The X system is chosen as $(R_x,R_y)$, the slow centre-of-mass coordinates. The MDE evolves monotonically during the relaxation process towards equilibrium. Although $N$ is not large and time-scale separation is not perfect, the MDE shows an almost monotonic increase.}
\label{fig:box-15}
\end{figure}

By the same numerical methods we calculate the time evolution of the MDE, which is shown in Fig.~\ref{fig:box-15} for the cases with $N=10$ and $15$ particles. Initially the particles are all in the left half of the box, and as time evolves the MDE increases until equilibrium is reached with some fluctuations due to the finite size. We have normalized $\Omega_Y(\Rv)$ so that the MDE regress to $0$. (This means that we have thrown away the Boltzmann entropy $\int \Omega_Y(\Rv) d\Rv$ which is a constant here.) As discussed in Sec.~\ref{sec:2nd-law}, time-scale separation ensures a monotonic MDE. Although systems with size $N=10$ are not usually thought capable of showing properly thermal behaviour, this is observed here.

\subsection{Synchronization in a Hamiltonian system}

We next study a mean-field model of $N$ coupled rotors on a circle \cite{AntoniRuffo1995,Dauxois2002,Campa2009,Latora1998}. This is a frictionless, noise-free Hamiltonian model that exhibits emergent, collective phase coherence (often described as ``synchronization'' in the Kuramoto sense). The phase variables are $\theta_i\in[0,2\pi)$ and their conjugate momenta are $p_i\in\mathbb{R}$, with Hamiltonian
\ba
H(\thetav,\pv)
=
\sum_{i=1}^N \frac{p_i^2}{2I}
+
\frac{K}{2N}\sum_{i,j=1}^N \left(1-\cos\left(\theta_i-\theta_j\right)\right),
\label{HMF-H}
\ea
where $I$ is the moment of inertia and $K>0$ is the coupling constant. The Kuramoto order parameter is
\ba
Re^{i\Psi} \equiv \frac{1}{N}\sum_{j=1}^N e^{i\theta_j},
\qquad
R\in[0,1].
\label{R-def}
\ea
Using the relation that
\ba
R^2 =\frac{1}{N^2}\sum_{i,j}\cos\left(\theta_i-\theta_j\right)
\label{cos-identity}
\ea
we can see that the potential energy depends only on $R$
\ba
V(\theta)=\frac{KN}{2}\left(1-R^2\right).
\label{V-R}
\ea
Thus $R$ is the correct macroscopic synchronization coordinate, whereas the global phase $\Psi$ is a Goldstone mode. 

According to the MDE formalism, we can define the conditional phase volume (CPV) at fixed total energy $E$ and order parameter $R$ as
\ba
\Omega_Y(R,E) \equiv
\int \prod_{i=1}^N d\theta_i\,dp_i\ 
\delta\left(E-H(\thetav,\pv)\right)\,
\delta\left(R-R(\thetav)\right).\nonumber\\
\label{OmegaRE-def}
\ea
The microcanonical equilibrium weight of $R$ is 
\ba
\rho^{\rm eq}(R)\propto \Omega_Y(R,E) = e^{S(R,E)},
\label{rhoR-prop}
\ea
where $S(R,E)$ can be viewed as an entropy landscape.

We denote the average energy of each rotor as $u = E/N$. In the limit $N\rightarrow \infty$, the system exhibits a phase transition at a critical energy $u_c={3K}/{4}$~\cite{Campa2009,AntoniRuffo1995,Latora1998}.
For $u<u_c$, the whole system is in a synchronous state. For $u>u_c$, the system moves randomly.  Here we study the cases with finite $N$. The order parameter $R$ has a distribution whose maximum lies at a $u$-dependent value $R^*(u)$. For large $u$, when there is no synchronization, $R^*$ is of order $1/\sqrt N$ as a result of random fluctuations, whereas in the synchronous phase it remains of order unity as $N\to\infty$. 

\begin{figure}[t]
\centering
\includegraphics[width=0.9\columnwidth]{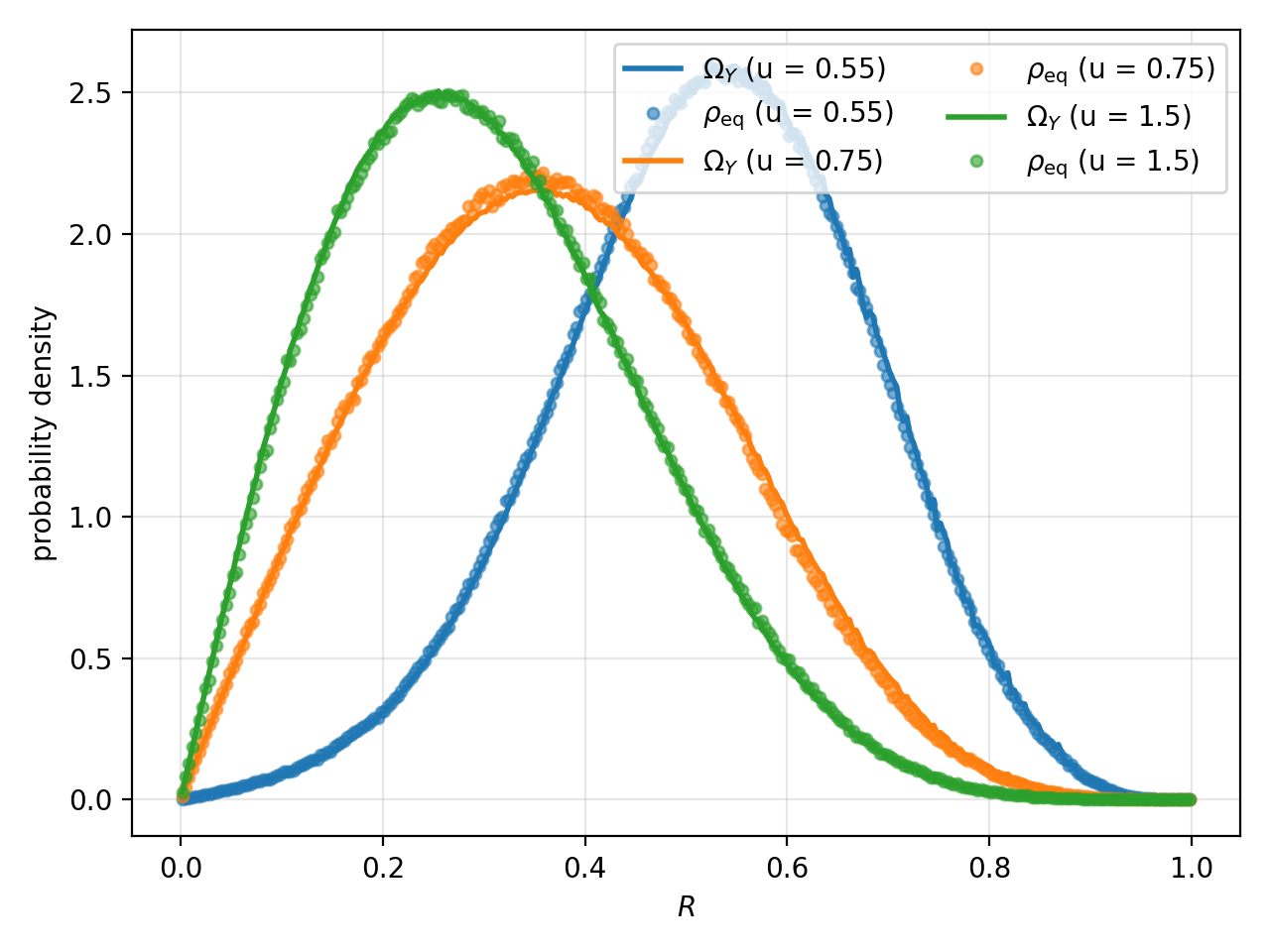}
\caption{Comparison of normalized conditional phase-space volume $\propto\Omega_Y(R)$ and the equilibrium distribution $\rho^{\rm eq}_X(R)$ for different average energies $u$ in the Hamiltonian synchronization model with $N=10$. The latter is obtained by analysing the long time visiting frequency.}
\label{fig:synchronization-eq}
\end{figure}

We numerically compute the normalized $\Omega_Y(R)$ and long-time distribution $\rho_{\rm eq}$ for $N=10$ and various average energies $u$. This is shown in Fig.~\ref{fig:synchronization-eq}. Here $\Omega_Y(R)$ is obtained by Monte Carlo sampling of angle configurations $\{\theta_i\}$, from which the order parameter $R=\left|\frac{1}{N}\sum_i e^{i\theta_i}\right|$ is recorded. Independently, the equilibrium distribution $p_{\rm eq}$ is obtained from microcanonical Hamiltonian molecular dynamics. Initial momenta are sampled and rescaled once at $t=0$ so that the total energy equals $E=Nu$, with the total momentum constrained to vanish. No further momentum rescaling is applied during the subsequent dynamics. Time evolution is performed using a velocity--Verlet integrator with timestep $\Delta t=0.01$, and energy conservation is monitored throughout the simulation. After a burn-in period, samples of $R(t)$ are collected from $\sim 500$ independent trajectories and aggregated to construct the empirical equilibrium density $\rho^{\rm eq}_X(R)$. The agreement between $\rho^{\rm eq}_X(R)$ and the normalized $\Omega_Y(R)$ is just as predicted.

\begin{figure}[t]
\centering
\includegraphics[width=0.9\columnwidth]{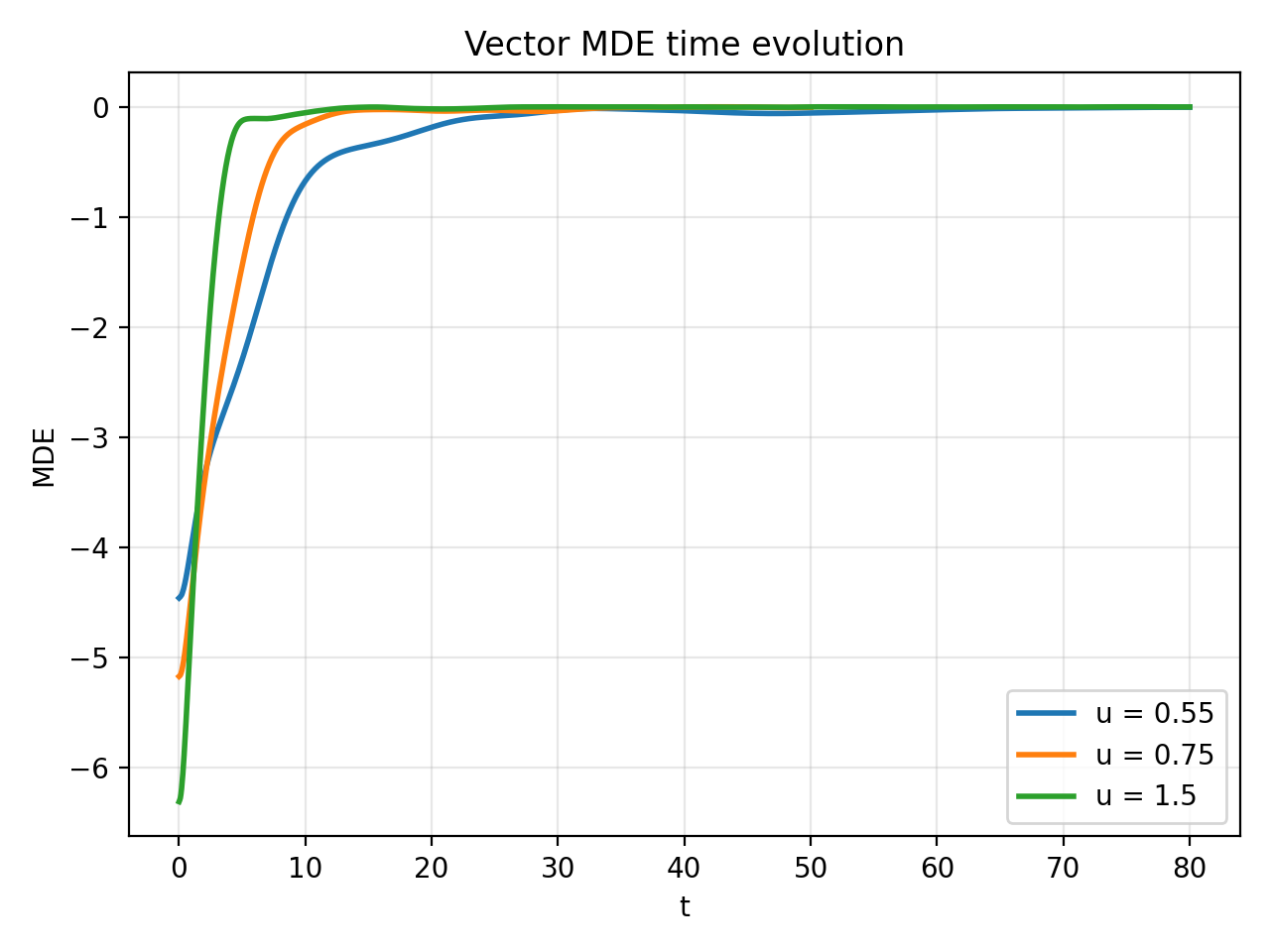}
\caption{The time evolution of the MDE (measured relative to its equilibrium value) for the Hamiltonian synchronization model with different average energy values $E/n = u$. The system size is $N=10$.}
\label{fig:synchronization-MDE}
\end{figure}

The above analysis discusses $R$ as though it were the only slow coordinate and the only member of X: notationally, we so far have implicitly chosen $\xv = R$. But both $R$ and $\Psi$ in Eq.~(\ref{R-def}) are actually slow variables. To secure time-scale separation in the MDE calculation we now include both in X, after first transforming them into 
\ba
R_x = \frac{1}{N}\sum_{i=1}^N \cos\theta_i,
\qquad
R_y = \frac{1}{N}\sum_{i=1}^N \sin\theta_i ,
\ea
and so choose $\xv = \Rv \equiv (R_x,R_y)$. Using the equations of motion $\dot\theta_i = p_i/I$, their exact time derivatives are
\ba
\dot R_x &=& -\frac{1}{NI}\sum_{i=1}^N p_i \sin\theta_i, \\
\dot R_y &=& \frac{1}{NI}\sum_{i=1}^N p_i \cos\theta_i .
\ea

At fixed energy density $u$, the momenta $p_i$ have finite variance of order $O(1)$.
Assuming weak correlations between different oscillators, the sums on the right-hand sides
satisfy a central-limit scaling,
\ba
\sum_{i=1}^N p_i \sin\theta_i = O(\sqrt{N}),
\qquad
\sum_{i=1}^N p_i \cos\theta_i = O(\sqrt{N}),
\ea
with vanishing means by rotational symmetry.
It follows that
$\dot R_x, \dot R_y$ are $\mathcal{O}(N^{-1/2})$, so that both components of $\Rv$ are indeed collective slow variables.

We have calculated the evolution for the MDE numerically for this system (with $\xv = \Rv$), with the results shown in Fig.~(\ref{fig:synchronization-MDE}) for three parameter values, $u=0.55, 0.75, 1.55$. As before, these entropies are measured relative to equilibrium values and all regress to zero. The system size is just  $N=10$. The initial angles are sampled from a Gaussian distribution centered at zero with a finite width. The initial momenta are drawn independently from a Gaussian distribution, shifted to enforce zero total momentum, and then uniformly rescaled so that the total energy satisfies $E = N u$. This defines a nonequilibrium initial condition on the microcanonical energy shell.
In all three cases the MDE increases toward its equilibrium value with moderate fluctuations about monotonicity. For practical purposes the MDE might indeed serve as a Lyapunov functional of the model, and should become one in the large $N$ limit.

\subsection{Zwanzig model as a non-ergodic paradigm}
\label{sec:beyond-ergodicity}

The construction developed above assumes that the unmonitored variables explore the full energy shell. This assumption is valid for ergodic systems. However, many finite Hamiltonian systems possess additional conserved quantities and are therefore not ergodic on the energy shell specified by $E$ alone.

To illustrate how the MDE framework extends to such situations, we consider the linear Zwanzig model~\cite{Zwanzig1973,Zwanzig2001},
\ba
H&=&\frac{P^2}{2M}+\frac{1}{2}M\omega_0^2Q^2 \nonumber
\\
&&+\sum_{i=1}^{N}\left[
\frac{p_i^2}{2m_i}
+\frac{1}{2}m_i\omega_i^2\left(q_i-\alpha_iQ\right)^2
\right].
\label{H-zwanzig-static}
\ea
Here $(Q,P)$ are the X-system variables and $(q_i,p_i)$ are the bath variables.

Since the Hamiltonian is quadratic, it can be diagonalized into independent normal modes,
\ba
H=
\sum_{\alpha=0}^{N}
\left[
\frac{1}{2}\Pi_\alpha^2
+\frac{1}{2}\Omega_\alpha^2R_\alpha^2
\right].
\label{H-normal}
\ea
The corresponding normal-mode energies
\ba
I_\alpha=
\frac{1}{2}\Pi_\alpha^2
+\frac{1}{2}\Omega_\alpha^2R_\alpha^2,
\qquad
\alpha=0,\ldots,N,
\label{normal-mode-energies}
\ea
are conserved.

As a result, a trajectory does not explore the full energy shell specified by $E$. Instead, the accessible phase-space region is restricted by the additional constraints $I_\alpha$. The conditional description must therefore be specified by $(X,E,I_1,\ldots,I_N).$
Accordingly, the CPV becomes
\ba
&&\Omega_Y^r(X,E,I_1,\ldots,I_N)
\nonumber\\
&=&
\int dY\,
\delta(E-H)
\prod_{\alpha=1}^{N}
\delta\left(
I_\alpha-\hat I_\alpha(X,Y)
\right).
\label{Omega-extra-conserved}
\ea
This generalized CPV counts only microstates compatible with both the total energy and the additional conserved quantities.
For the present example we choose the selected variable as $Q$ and integrate out all remaining variables, including the momentum $P$. 

\begin{figure}
\includegraphics[width=0.45\textwidth]{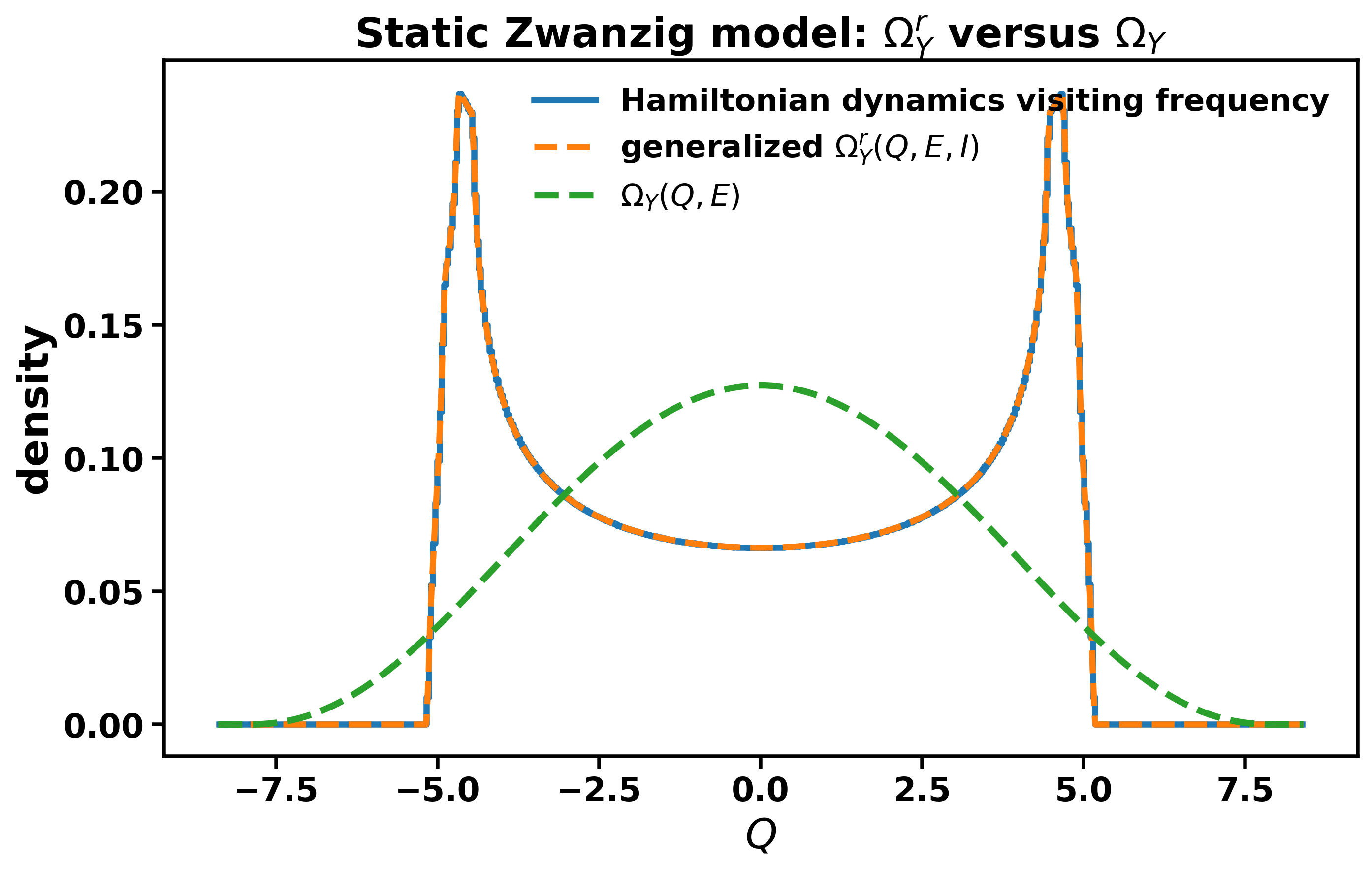}
\caption{
Static $N=3$ linear Zwanzig model. The blue curve is the exact long-time visiting frequency of $Q$ obtained from Hamiltonian dynamics. The orange curve is the normalized distribution predicted by the generalized CPV $\Omega_Y^r(Q,E,I_1,\ldots,I_N)$. The green dashed curve is the prediction based on the original CPV $\Omega_Y(Q,E)$. The generalized CPV correctly reproduces the observed distribution, whereas the original CPV fails because it ignores the additional conserved quantities.
}
\label{fig:zwanzig-static}
\end{figure}

For the numerical test, we use $N=3$, $M=1$, $m_i=1$, $\omega_0=0.5$, and $E=8$. The coupling scale is $0.18$. The additional conserved quantities are fixed as
\ba
I_1=1.20,\qquad I_2=1.80,\qquad I_3=2.10.
\ea
Hamiltonian trajectories are generated with these fixed values of $(E,I_1,I_2,I_3)$ and random initial phases. The resulting long-time visiting frequency of $Q$ is compared with the normalized distribution predicted by $\Omega_Y^r(Q,E,I_1,I_2,I_3)$. We also compare with the original CPV, $\Omega_Y(Q,E)$, which retains only the total-energy constraint. Figure~\ref{fig:zwanzig-static} shows that $\Omega_Y^r$ reproduces the observed visiting frequency almost perfectly, whereas $\Omega_Y$ gives a qualitatively incorrect prediction. The conserved quantities $I_\alpha$ confine the dynamics to invariant manifolds that occupy only a small fraction of the full energy shell. Consequently, many values of $Q$ allowed by $\Omega_Y(Q,E)$ are never visited dynamically, leading to the strongly non-Gaussian distribution observed in the figure.

We emphasize that this example is intended to test the generalized CPV $\Omega_Y^r$ through the long-time visiting frequency of the selected variable. It is not intended as a test of monotonic entropy production. Because the number of bath oscillators is small and the resulting dynamics is quasi-periodic rather than mixing, the assumptions required for monotonic increase of the MDE are not expected to hold in this example.

\vspace{5mm}
\section{Choice of Reduced Coordinates}
\label{sec:information}

The MDE developed in this paper defines a nonequilibrium entropy directly from classical mechanics. At fixed energy $E_{XY} = E$  its equilibrium limit is $S_B(E)=\log \int d\xv\,\Omega_Y(\xv,E)$. This is the usual Boltzmann entropy of the full XY system. As such, for systems of similar particles such as the $N$ particles in a hard-walled box considered in Sec.~\ref{sec:box}, it lacks the celebrated $1/N!$ factor introduced by Gibbs to render the equilibrium entropy extensive in the thermodynamic limit of large $N$.

While clearly established in equilibrium, the dynamical status of the Gibbs factor is not always clear~\cite{Cates2015}. This can be clarified with the aid of the MDE, and we do this in the current Section. Specifically, in a classical Hamiltonian setting, the Gibbs factor encodes the fact that not only the thermodynamic observables, but also the Hamiltonian itself, are invariant under permutation of particle identities. The dynamical validity of the Gibbs factor in that case is established explicitly in \eqref{Omega-Yprime} below.

The distinction between permutation symmetry and indistinguishability is significant because there are many systems, such as colloids, in which strictly classical particles are {\em a priori} physically distinguishable (for instance, each colloid has a different number of atoms) but are not distinguished in practice \cite{Sethna2021,Cates2015,Meng2010,Arkus2009,Perry2015}. While the $N!$ is often attributed in textbooks to quantum indistinguishability, this is not relevant in such cases. It is also not the argument given by Gibbs whose explanation of the $N!$ was, of course, entirely classical. What is really at issue is not whether classical particles are distinguishable, but whether the chosen thermodynamic variables actually distinguish particle labels~\cite{Sethna2006}. 

In an MDE setting it might thus be tempting to include particle labels among the unmonitored degrees of freedom Y. However, this temptation should be resisted because an assumption of fast bath relaxation is required for our proof above (Sec.~\ref{sec:2nd-law}) of the second law for the MDE. Yet it is found experimentally that many colloidal systems do fully obey all the laws of thermodynamics even on timescales such that --- as discussed in detail in~\cite{Cates2015} --- their particle labels barely relax.

Instead, the MDE allows us to separate two questions that are often conflated: what variables define the thermodynamic description, and whether the system is actually in equilibrium with respect to those variables.

\subsection{Generalized variables}
\label{sec:general-x}
To make the above point precise, we generalize the MDE construction so that the selected variables $\xv$ need not be a subset of canonical coordinates as assumed in previous sections. (For example $\xv$ could represent the empirical density, rather than the particle coordinates, of a set of colloids within a thermal solvent Y.)

The canonical assumption has been convenient so far, because the Liouville measure then factorizes as $d\xv d\yv$.
More generally, one may choose a set of selected variables $\zv(\xv,\yv)$ as arbitrary functions of the canonical microscopic coordinates $(\xv,\yv)$. The total Hamiltonian is still denoted by $H_{XY}(\xv,\yv)$, and the underlying measure remains the Liouville measure $d\xv d\yv$. The reduced distribution and conditional phase-space volume are then defined as:
\ba
\rho_{ZE}(\zv,E)
&=&
\int d\xv d\yv\, \rho_{XY}(\xv,\yv)
\delta\left(\zv - \zv(\xv,\yv)\right) \nonumber\\
&&\delta\left(E - H_{XY}(\xv,\yv)\right),
\label{rhoZE-general}
\\
\Omega_Z(\tilde\xv,E)
&=&
\int d\xv d\yv\,
\delta\left(\zv - \zv(\xv,\yv)\right)
\delta\left(E - H_{XY}(\xv,\yv)\right).\nonumber\\
\label{OmegaZ-general}
\ea
The corresponding MDE is
\ba
S^{\rm md}(\zv,E,t)
=
-\log \rho_{ZE}(\zv,E,t)
+
\log \Omega_Z(\zv,E).
\label{Smd-general-Z}
\ea
As mentioned, $\zv$ is now an arbitrary set of retained coordinates. Different choices of $\zv(\xv,\yv)$  correspond to different thermodynamic descriptions of the same Hamiltonian system whose dynamics is not itself affected by that choice. For notational simplicity, hereafter we drop the $\zv$ notation and understand that $\xv$ is an arbitrary rather than canonical subset of coordinates.

\subsection{Permutation sectors and thermodynamic CPV}
\label{sec:gibbs-factor}
Consider a system of $N$ identical classical particles. The microscopic phase space is `labelled' (each particle has a separate coordinate), but the Hamiltonian is invariant under particle permutations. The variables probed by ordinary thermodynamic measurements, such as empirical densities or coarse-grained density fields, are also invariant under particle relabelling. 

For such systems it is useful to decompose the microscopic degrees of freedom as
\ba
(\xv,\yv)\longrightarrow (\xv,\yv',\pi),
\label{XY-pi-decomp}
\ea
where $\xv\in$ X denotes the retained thermodynamic variables, $\yv'\in$ Y$'$ denotes the genuinely fast unobserved variables, and $\pi$ is a label specifying which permutation sector the system occupies. Both $\xv$ and $\yv$ are invariant under permutations. Equivalently, they are both functionals of the empirical particle density, whereas $\pi$ specifies one of a set of $N!$ discrete maps that convert the empirical density to a list of $N$ particle coordinates. 

Note that there are many way to {\em parameterize} the set of maps $\{\pi\}$. One is to follow, up to the present time $t$, the historical Hamiltonian dynamics from initial conditions that differ only in particle labels. Another, time-local, choice is to create a list of particle coordinates in some algorithmic order (such as in order of their radial distance from a given origin, which is unique with probability one) such that $\pi$ selects one of the $N!$ ways in which particle labels can then be assigned to this ordered list.

Using such a construction, X contains only the slow variables of thermodynamic interest, while Y$'$ are the degrees of freedom whose rapid equilibration can be exploited to derive thermodynamics via the MDE. Importantly, the permutation label $\pi$ is often not equilibrated, hence also a slow variable, and hence not in Y$'$.  

For example, in a colloidal crystal the identity of the colloids occupying particular lattice sites may remain effectively fixed throughout the observation time, whereas in a colloidal fluid phase there is significant, if still sub-ergodic, exploration of the permutation sectors~\cite{Cates2015}. This nonergodicity, even if phase-dependent as in this case, does not prevent emergence of thermodynamics for the X subsystem, because this only requires equilibration of the unmonitored variables that {\em can influence} the dynamics of $\xv \in$ X. Since $H(\xv,\yv',\pi)$ is invariant under permutations, $\pi$ is not among those variables.

Given permutation symmetry and the resulting dynamical invisibility of $\pi$, the conditional phase-space volume governing the evolution in X can be chosen as
\ba
\Omega_{Y'}(\xv,\pi,E)
=
\int dY'\,
\delta\left(E-H(\xv,\yv',\pi)\right).
\label{Omega-Yprime}
\ea
This captures the fact that relaxation is governed by the fast variables $\yv'$ rather than the slow permutation labels $\pi$. 
In \eqref{Omega-Yprime} the choice of sector $\pi$ is arbitrary. Moreover, since $H$ does not depend on $\pi$, this $\Omega_{Y'}$ is independent of the $\pi$ label which may then be dropped, giving
\ba
\Omega_{Y'}(\xv,\pi,E)
=
\Omega_{Y'}(\xv,E) = \Omega_{Y}(\xv,E)/N!.
\label{Omega-pi-independence}
\ea
Equation~\eqref{Omega-pi-independence} justifies use of the Gibbs factor under both equilibrium and non-equilibrium conditions whenever $H(\xv,\yv',\pi)$ has permutation symmetry. Calculation of the MDE then proceeds as before, but with $\Omega_{Y'}$ replacing $\Omega_Y$ as the CPV in Eq.~\eqref{def-mde-0}.

The MDE formulation also clarifies the entropy of mixing. Suppose that the particles are assigned observable colours. Even if the microscopic dynamics do not change (as holds for a colour-blind Hamiltonian), the selected variables $\xv \in$ X do change. Part of the permutation-sector information that was in $\pi$ and previously ignored now becomes thermodynamically observable and must be moved into X. For a mixture containing $N_r$ red particles and $N_b$ blue particles, the corresponding Gibbs divisor in \eqref{Omega-pi-independence} is therefore changed from $N!$ to
$
{N_r!N_b!}
$
which, through the term $\log \Omega_{Y'}$, further alters $S^{\rm md}$ in Eq.~\eqref{def-mde-0} by the standard entropy of mixing, $\Delta S^{\rm mix} = \log(N!/N_r!N_b!)$. Importantly, while equilibration of permutations within each colour is irrelevant as before, the $\Delta S^{\rm mix}$ term is time-independent, and present whether or not the colour labels are themselves equilibrated. However, since those labels now reside in X, they can influence the time evolution of $S^{\rm md}$ directly.

This example illustrates a central theme of the MDE framework: thermodynamic equilibrium is defined relative to the selected variables, not to the full microscopic phase space. The Gibbs factor and the entropy of mixing arise because different choices of selected variables determine which permutation-sector information is regarded as thermodynamically relevant.

To summarize the above arguments: the dynamical exploration of the permutation sectors $\pi$, whether fast or slow, makes no contribution to the time evolution of the MDE in any system whose thermodynamic coordinates $\xv\in$ X are independent of $\pi$. 

Crucially, any attempt to formulate the thermodynamical entropy in dynamical terms, without accepting that point, can lead to major paradoxes. For example, if one assigns an entropy that varies according to how well $\pi$ is equilibrated, one finds that colloidal crystals (localized particles) suffer an entropy penalty relative to the fluid (mobile particles) that is large enough to always destabilize the crystalline phase, contrary to observation~\cite{Cates2015}. 

The arguments of this Section resolve such paradoxes quantitatively, not just in equilibrium but for all observation times. So long as both the Hamiltonian and the chosen thermodynamic variables are invariant under particle permutations, it is immaterial to what extent the sector label $\pi$ equilibrates on observable timescales. Consequently, our previous discussions (Sec.~\ref{sec:2nd-law}) concerning autonomy, time-scale separation, and the second law remain fully applicable to the dynamics of the X subsystem even in cases, such as colloids, where particle permutations are not equilibrated on practical time scales. 

Of course, the same is {\em not} true if the $\pi$-invariance of the Hamiltonian, that we depended upon above, is operationally absent. This applies for colloids with appreciable polydispersity in size (for example). Here the slow dynamics of size segregation can prevent equilibration even of variables that do not themselves distinguish among particle sizes. Such variables include the usual order parameters for colloidal crystallinity; for broad size distributions, crystallization can {\em only} proceed when accompanied by size fractionation~\cite{Bartlett2000}. A thermodynamically extensive MDE for continuous polydispersity can be constructed, just as in equilibrium, by grouping the particle size into $\alpha = 1...m$ bins, applying the Gibbs factor $\Pi_{\alpha}N_\alpha!$, and then taking $m\to\infty$ in a volume-independent manner~\cite{Sollich2001}.

\if{
\mec{{\bf I think the following stuff is no longer needed:}
Just as in \eqref{def-mde-minus}, the thermodynamic entropy associated with the variables $\xv$ depends only on $\rho_X$ and $E$:
\ba
S^{\rm md}_{\rm th}
=
S_X
+
\left\langle
\log \Omega_{Y'}(X,E)
\right\rangle .
\label{Smd-th}
\ea

The Gibbs factor follows directly from this separation. If the sector label is not separated explicitly and is instead included among the unmonitored variables, the corresponding conditional phase-space volume counts all permutation sectors:
\ba
\Omega_Y(X,E)
=
\sum_\pi \Omega_{Y'}(X,\pi,E).
\label{OmegaY-sum-pi}
\ea
Equation (\ref{Omega-pi-independence}) gives
\ba
\Omega_Y(X,E)
=
N!\Omega_{Y'}(X,E).
\label{OmegaY-factorial}
\ea
Since thermodynamic relaxation is governed by the fast variables Y' rather than the permutation-sector label $\pi$, the thermodynamic CPV is identified with $\Omega_{Y'}$ rather than $\Omega_Y$. Equivalently,
\ba
\Omega_{Y'}(X,E)
=
\frac{1}{N!}\Omega_Y(X,E).
\label{Omega-gibbs-factor}
\ea
Thus the Gibbs factor appears when the non-ergodic permutation-sector contribution is excluded from the thermodynamic conditional phase-space volume.

The same factor appears in the equilibrium limit. The thermodynamic phase-space volume is
\ba
\Omega^{\rm th}(E)
=
\int dX\,\Omega_{Y'}(X,E)
=
\frac{1}{N!}
\int dX\,\Omega_Y(X,E),
\label{Omega-total-th}
\ea
and therefore the Boltzmann entropy
\ba
S_B(E)
=
\log \Omega^{\rm th}(E)
\label{Boltzmann-gibbs}
\ea
contains the familiar correction $-\log N!$.}

}\fi

\vspace{5mm}

\section{Comparison with other approaches}
\label{mec_others}

For systems confined to a fixed energy shell as considered here (but generalized in Ref.~\cite{ding2026MDE2}), the MDE admits a transparent alternative decomposition. Using
\ba
\rho_X^{\rm eq}(\xv,E)
= \frac{\Omega_Y(\xv,E)}{\int d\xv\,\Omega_Y(\xv,E)},
\label{mde-eq-dist-compare}
\ea
the MDE can be written as 
\ba
S^{\rm md}_{XY}(E,t)
= S_B(E)
- D\left[\rho_X(t)\middle\|\rho_X^{\rm eq}(E)\right],
\label{mde-boltz-kl}
\ea
where the first term is a Boltzmann-like equilibrium entropy, $S_B(E)=\log \int d\xv\,\Omega_Y(\xv,E)$, and the second is
a Kullback-Leibler relative entropy:
\ba
D\left[\rho_X(t)\middle\|\rho_X^{\rm eq}(E)\right]
= \int d\xv\,\rho_X(\xv,t)
\log\frac{\rho_X(\xv,t)}{\rho_X^{\rm eq}(\xv,E)}.
\label{mde-kl-def-compare}
\ea
The latter measures the distance from equilibrium and therefore evolves towards zero under relaxation. For time-independent Hamiltonians as considered here, the equilibrium thermodynamic entropy $S_B(E)$ is a constant that does not contribute to the entropy production or to irreversibility in general. However, for time-dependent Hamiltonians, the corresponding, equilibrium-like term becomes dynamical and plays a central role~\cite{ding2026MDE2}.

The above decomposition clarifies the differences between our MDE and related constructions known in the literature. One commonly used ansatz for a microscopic and dynamical definition of entropy is
\ba
S_X = S_X^G - \beta \int d\xv\,\rho_X(\xv,t) H_X(\xv),
\label{alt-beta-entropy}
\ea
where \(H_X\) may denote either the bare subsystem Hamiltonian or an effective Hamiltonian such as the Hamiltonian of mean force~\cite{jarzynski2004,Seifert2012,Campisi2009,Talkner2020}. Such a definition depends explicitly on \(\beta\), which is not an intrinsic property of subsystem X ({\em i.e.}, unlike our MDE, this expression is {\em not} a functional of the marginal $\rho_x(\xv)$ only). 

Maximization of \eqref{alt-beta-entropy} gives a Gibbsian form \(\rho_X^{\rm eq}\propto e^{-\beta H_X}\). This is an approximation, valid only in the large-bath limit, whereas for finite isolated XY systems the correct equilibrium is instead given by Eq.~(\ref{mde-eq-dist-compare})~\cite{Gross1997,Gross2001}. This point is further illustrated by the Zwanzig model in Sec.~\ref{sec:beyond-ergodicity}, where the equilibrium distribution is not determined by $\Omega_Y$ itself but by the generalized CPV $\Omega_Y^r$. Since the Gibbs distribution arises only as a further approximation to $\Omega_Y$, its range of validity is correspondingly more limited.

The appearance of a relative-entropy structure in Eq.~(\ref{mde-boltz-kl}) is noteworthy because KL divergences have often been used to characterize relaxation toward equilibrium, {\em e.g.}, by serving as a nonequilibrium free energy~\cite{Spohn1978,Qian2001}. Nevertheless, the physical interpretation of such relative-entropy constructions remains the subject of ongoing discussion, particularly away from equilibrium and beyond detailed-balance settings~\cite{Maes2012}. 

Although true for a time-independent Hamiltonian and a fixed energy shell, it is not true that our MDE only ever differs from the Boltzmann entropy by a KL relative entropy quantifiying the distance from the equilibrium distribution.
As shown in Eq.~(\ref{mde-kl-form}) of Paper II~\cite{ding2026MDE2}, the general nonequilibrium entropy contains additional contributions associated with the energy distribution itself and the resulting thermodynamic structure. These contributions are absent from a purely relative-entropy description. Therefore, while KL divergences provide a useful characterization of relaxation, they do not by themselves account for the microscopic multiplicity encoded by the CPV, and hence do not furnish a complete microscopic dynamical entropy for Hamiltonian systems.

There are further alternative theories linking entropy production to the time-dependent correlation between the X and Y subsystems~\cite{Esposito2010,Dolatkhah2020}. However, the entropies involved depend on the bath details. Hence they do not define an intrinsic dynamical entropy that is a functional only of the total energy $E$ and the marginal phase-space density $\rho_X(\xv)$ of subsystem X alone, as we have done. We consider that to be a core requirement of the MDE which is intended as a microscopically defined quantification of the thermodynamic entropy, which itself does not depend on bath details.

Finally, stochastic-thermodynamic definitions based on forward/backward path probability ratios provide a key characterization of irreversibility for coarse-grained stochastic dynamics~\cite{Seifert2012}. Such constructions rely on the existence of a closed, autonomous description at the level of the reduced variables X (such as a Fokker-Planck or Langevin equation for these), so that path probabilities for X are well defined and satisfy appropriate consistency conditions. In fully general Hamiltonian systems, however, the reduced dynamics of X is not autonomous, and a direct application of such path-based definitions is therefore not straightforward. How to consistently extend the stochastic thermodynamics approach to Hamiltonian systems, by identifying the appropriate variables and path measures, will be addressed in Paper II~\cite{ding2026MDE2}.

\vspace{5mm}

\section{Concluding remarks}
Any candidate for a microscopic dynamical entropy describing a subsystem X within a finite Hamiltonian system XY should satisfy several basic requirements, as follows. It should reproduce the correct equilibrium distribution of X when maximized; it should be expressed in terms of the state of X, as encoded in its marginal density $\rho_X(\xv)$, without explicit dependence on microscopic details of the bath Y; in the large-bath limit, it should reproduce the standard entropy balance $\Delta S_{\rm tot}=\Delta S_X-\beta Q$, where $Q$ is the heat absorbed by X; and, under appropriate dynamical conditions such as autonomy or mixing, it should be non-decreasing while allowing exceptions under controlled reversals of momenta such as spin echo. 

For conceptual clarity, it is also desirable that such an entropy should not explicitly involve the bath temperature $T$, which, from the viewpoint of Hamiltonian dynamics, is itself an emergent macroscopic quantity. A single bath temperature is well defined only under additional thermodynamic assumptions, such as a macroscopic bath close to equilibrium; these assumptions need not hold for finite baths or for far-from-equilibrium initial states of the full XY system. (The latter include, for example, the case where Y is made up of two sub-baths at different temperatures.) As we have shown, the MDE naturally satisfies these requirements. The MDE satisfies these requirements. Importantly, it is not postulated as an entropy functional, but emerges directly from the reduction of the full description $(\xv,\yv)$ to a reduced description $(\xv,E)$, where $E=H_{XY}$ is the total energy of the XY system.

A central concept in our MDE framework is the conditional phase-space volume (CPV), $\Omega_Y$ (or for identical particles $\Omega_{Y'}$), which provides the fundamental statistical quantity governing the reduced description of Hamiltonian systems. Through Eq.~(\ref{mec_micro}), $\Omega_Y$ determines the equilibrium distribution $\rho_X^{\rm eq}(\xv)$ of the observed variables X. This result follows directly from the dynamical visitation properties of trajectories and does not rely on an entropy postulate, coarse-graining procedure, or stochastic assumptions. In this sense, $\Omega_Y$ provides a purely dynamical characterization of equilibrium at the level of the X subsystem.

The MDE emerges naturally from $\Omega_Y$ and inherits its statistical and thermodynamic significance. It arises directly from the multiplicity of microscopic states compatible with a given reduced description. By construction, it is maximized by $\rho_X^{\rm eq}$ and remains fully compatible with time-reversible Hamiltonian dynamics. Since the equilibrium distribution itself is determined by $\Omega_Y$, these requirements place strong constraints on admissible entropy functionals. Accordingly, the CPV serves as the central geometric object underlying the entire MDE framework. 

In general nonequilibrium situations the geometric structure encoded in the MDE constrains, but does not determine, the detailed time evolution. However, when there is a separation of time scales between X and Y, the quantity $\log \Omega_Y(\xv,E)$ reduces to the generalized potential identified in Refs.~\cite{ding2020,ding2022,ding2022-2}. Similar effective potentials arising from the elimination of fast variables have long been known in projection-operator and coarse-graining approaches~\cite{Zwanzig1961,mori1965,Grabert1982,Zwanzig2001}. Together with mobility function(al)s that depend on the underlying Hamiltonian, they govern the effective Langevin or Fokker--Planck dynamics of X. In this regime, the MDE therefore connects directly to stochastic descriptions while providing a microscopic geometric interpretation of the resulting effective potential.

More broadly, our MDE framework clarifies the microscopic origin of irreversibility. The very notion of a heat bath, central to thermodynamics, implicitly introduces an arrow of time: when a system X is first brought into contact with a bath Y, the combined system is initially weakly correlated, whereas subsequent evolution generates strong system--bath correlations. A consistent entropy must therefore encode this asymmetry through its choice of coarse-graining. By incorporating the bath degrees of freedom via the CPV, $\Omega_Y$, the MDE provides such a construction, capturing both equilibrium structure and the emergence of irreversibility within the setting of Hamiltonian dynamics.

From this perspective, the MDE is not merely one possible definition among many, although it may coincide with alternative constructions in certain limits (see Sec.~\ref{mec_others}). Rather, it emerges as a distinctively natural microscopic entropy functional that reconciles thermodynamics with Hamiltonian dynamics for both large and small systems of XY type. For mixing systems with autonomy and time-scale separation, the MDE quantifies monotonic thermal relaxation, while allowing exceptions in situations (such as echo experiments) where those conditions fail. In this sense, the Microscopic Dynamical Entropy of Eq.~(\ref{def-mde-0}) provides a unified microscopic formulation of entropy for Hamiltonian systems.

Finally, it is natural to consider extensions of this framework to quantum systems. For a closed system with Hamiltonian $\hat H_{XY}$, our classical energy-shell construction is replaced by the microcanonical projector
\ba
\hat\Delta = \delta_\epsilon(E-\hat H_{XY})
= \sum_{k:|E_k-E|\le\epsilon}|E_k\rangle\langle E_k|.
\ea
This suggests a quantum analogue of the MDE,
\ba
S^{\rm MD}_{XY} = S^{\rm vN}_X + \Tr_X \hat \rho_X\log \hat \Omega_Y, 
\qquad
\hat \Omega_Y = \Tr_Y \hat \Delta,
\ea
where $\hat\rho_X = \Tr_Y \hat\rho_{XY}$. This construction provides a candidate microscopic entropy for quantum systems, which appears to be distinct from previous proposals~\cite{Talkner2020,Landi2021}, and which we plan to explore in future work.

{\em Acknowledgements:} We thank Xiangjun Xing for important discussions in the early stages of this work, Michael te Vrugt for a most interesting seminar on irreversibility problems, Christopher Jarzynski for a critical reading of a much earlier version of the manuscript~\cite{ding2025}, and Jun Wu and Hanchun Wang for their help in our simulation studies.

\bibliographystyle{unsrt}
\bibliography{reference}

\end{document}